\documentstyle[epsf]{mn}

\def\Refsec#1{Section~\ref{sec:#1}}
\def\Refapn#1{Appendix~\ref{apn:#1}}
\def\Refeq#1{equation~(\ref{eq:#1})}
\def\refeq#1{(\ref{eq:#1})}
\def\Reffig#1{Fig.~\ref{fig:#1}}

\def\Peacock81{Peacock \shortcite{Pe81}}
\def\KS87{Kirk \& Schneider \shortcite{KS87}}
\def\EJR90{Ellison et al. \shortcite{EJR90}}
\def\Kirk88{Kirk \shortcite{Ki88}}
\def\Bell78{Bell \shortcite{Be78}}

\def\subu#1{{#1_{\rm u}}}
\def\subd#1{{#1_{\rm d}}}
\def\subf#1{{#1_{\rm f}}}
\def\subrel#1{{#1_{\rm rel}}}
\def\subp#1{{#1_{+}}}
\def\subm#1{{#1_{-}}}
\def\subD#1{{#1_{\rm D}}}
\def\subDz#1{{#1_{\rm D0}}}
\def\subud#1{{#1_{\rm ud}}}
\def\subdu#1{{#1_{\rm du}}}
\def\subuw#1{{#1_{\rm UW}}}
\def\subdw#1{{#1_{\rm DW}}}

\def\Pmu{P_{\mu'}}
\def\Vf{\subf{V}}
\def\Gf{\subf{\Gamma}}
\def\Dx{{\Delta x}}
\def\Dt{{\Delta t}}
\def\lmd{{\lambda^*}}
\def\Lp{L_{+}}
\def\Lm{L_{-}}

\def\nN{{n_N}}
\def\nNz{{n_N^{(0)}}}
\def\th0{{\theta_0}}

\def\Pabs{P_{\rm abs}}
\def\Pret{P_{\rm R}}
\def\Pdw{P_{\rm DW}}
\def\Puw{P_{\rm UW}}

\def\lmdz{{\lambda^*_0}}
\def\LD{\subD{L}}
\def\LDz{\subDz{L}}
\def\Rp{R_{+}}
\def\Rm{R_{-}}
\def\Pm{\subuw{P}}
\def\Pp{\subdw{P}}
\def\Np{\subp{N}}
\def\Nm{\subm{N}}
\def\gp{\subp{g}}
\def\gm{\subm{g}}
\def\hm{\subm{h}}

\def\Vd{\subd{V}}
\def\Gd{\subd{\Gamma}}
\def\dd{\subd{\nu}}
\def\mud{\mu^{\rm (d)}}
\def\mudz{\mu_0^{\rm (d)}}
\def\mudo{\mu_1^{\rm (d)}}
\def\lmdd{{\subd{\lambda}^*}}
\def\Pdwd{P_{\rm DW}^{\rm (d)}}

\def\Pretd{P_{\rm R}^{\rm (d)}}
\def\Vu{\subu{V}}
\def\Gu{\subu{\Gamma}}
\def\du{\subu{\nu}}
\def\muu{\mu^{\rm (u)}}
\def\muuz{\mu_0^{\rm (u)}}
\def\muuo{\mu_1^{\rm (u)}}

\def\Puwu{P_{\rm UW}^{\rm (u)}}
\def\Vrel{\subrel{V}}
\def\Grel{\subrel{\Gamma}}

\def\Vs{V_{\rm s}}
\def\pdu{\subdu{\Phi}}
\def\pud{\subud{\Phi}}
\def\PRET{{\langle P_{\rm R}\rangle}}
\def\lnd{{\langle\ln\delta\rangle}}
\def\lndud{\subud{\langle\ln\delta\rangle}}
\def\lnddu{\subdu{\langle\ln\delta\rangle}}

\def\muuud{\mu_{\rm ud}^{\rm (u)}}
\def\muddu{\mu_{\rm du}^{\rm (d)}}

\def\mudud{\mu_{\rm ud}^{\rm (d)}}

\def\dilog#1{{\rm dilog} \left( #1 \right)}
\def\LDu{L_{\rm D}^{\rm (u)}}
\def\LDd{L_{\rm D}^{\rm (d)}}

\def\Rpu{R_{+}^{\rm (u)}}

\def\chiz{{\chi_0}}
\def\chione{{\chi_1}}
\def\chiu{{\chi_{\rm u}}}
\def\chid{{\chi_{\rm d}}}
\def\Rc{\tilde{R}}

\def\nabs {n_{\rm pass}}

\begin{document}

\title[Random walk theory of shock acceleration]
{Application of random walk theory\\
to the first order Fermi acceleration in shock waves}

\author[T. N. Kato and F. Takahara]
{T. N. Kato and F. Takahara\\
Department of Earth and Space Science,
Graduate School of Science,\\
Osaka University, 
Machikaneyama 1-1, Toyonaka, Osaka 560-0043, Japan}

\maketitle

\begin{abstract}
We formulate the first order Fermi acceleration in shock waves 
in terms of the random walk theory. 
The formulation is applicable to any value of the shock speed
and the particle speed,
in particular,
to the acceleration in relativistic shocks and
to the injection problem, where
the particle speed is comparable to the fluid speed,
as long as large angle scattering is suitable for
the scattering process of particles. 
We first show that the trajectory of a particle suffering from 
large angle scattering can be treated as 
a random walk in a moving medium with an absorbing boundary 
(e.g., the shock front). 
We derive an integral equation to determine the density of scattering points
of the random walk,
and by solving it approximately
we obtain approximate solutions of
the probability density of pitch angle at and the return probability 
after the shock crossing in analytic form.
These approximate solutions
include corrections of several non-diffusive effects
to the conventional diffusion approximation
and we show that
they agree well with the Monte Carlo results
for isotropic scattering model
for any shock speed and particle speed.  
When we neglect effects of `a few step return',
we obtain `the multi-step approximation'
which includes only the effect of finite mean free path
and which is equivalent
to `the relativistic diffusion approximation' used by Peacock (1981)
if the correct diffusion length is used in his expression.
We find that
the multi-step approximation
is not appropriate to describe the probability densities of individual particles
for relativistic shocks,
but that
the pitch angle distribution at the shock front in steady state
is in practice quite well approximated by that given by
the multi-step approximation
because the effects of finite mean free path and a few step return
compensate each other
when averaged over pitch angle distribution. 
Finally, we give an analytical expression of the spectral index 
of accelerated particles in parallel shocks
valid for arbitrary shock speed
using this approximation. 
\end{abstract}

\begin{keywords}
acceleration of particles -- methods:analytical -- shock waves -- cosmic rays.
\end{keywords}

\section{Introduction}
\label{sec:introduction}
It is now widely accepted that the first order Fermi acceleration,  
which is driven in various astrophysical shocks, 
provides a factory of non-thermal energetic particles.
For example, recent observations discovered synchrotron X-rays
from energetic electrons with energies above $10^{14}$eV
in the shell of the supernova remnant 1006 \cite{KPG95}. 
Since the basic theory was proposed more than 20 years ago by 
\Bell78, Krymski (1977), Axford et al. (1977), and 
Blandford and Ostriker (1978), many papers have dealt with 
fundamental mechanisms of shock acceleration.
The mechanism has been applied to many situations ranging 
from interplanetary shocks to relativistic shocks
in active galactic nuclei or gamma-ray bursts 
(see for review Blandford \& Eichler 1987 and Kirk \& Duffy 1999). 
Although most of the recent theoretical interests 
have been concentrated on non-linear problems where the shock 
structure is modified by accelerated particles
(Ellison, Baring \& Jones 1996,
Berezhko \& Ellison 1999), 
some of the linear problems in the test particle approximation 
still remain to be clarified and need 
more close examinations, especially for 
relativistic shocks, oblique shocks and injection of 
seed particles.

The first order Fermi acceleration in shock waves 
with test particle approximation
is the most basic problem of the shock acceleration.
As a particle gains energy
whenever the particle repeats shock crossing and recrossing 
between the upstream and downstream of the shock front,
it is essential to calculate the average energy gain per this cycle
and the number distribution of the repeated cycles 
before the particles escape to far downstream. 
These are determined by the shock speed,
the compression ratio of the shock, 
the pitch angle distribution of particles at the shock crossing 
and the return probability of particles, i.e., the probability of 
recrossing the shock front for particles that cross the front 
from the upstream to the downstream. 
As is well known, the energy spectrum of particles accelerated 
by this mechanism takes a power law form, 
$F(E)dE \propto E^{-\sigma}dE$. 
For the non-relativistic shocks, the conventional diffusion 
approximation can be used to obtain the universal power 
law spectrum with an index of $\sigma = (r+2)/(r-1)$,
where $r$ is the compression ratio of the shock (Bell 1978a).
However, for relativistic shocks, oblique shocks and acceleration 
of low energy particles,
in which the fluid speed is not negligible
compared with the particle velocity,
anisotropy of particle distribution at the 
shock front becomes large and the diffusion approximation is not 
relied on. Thus, one should develop more sophisticated treatments 
beyond the diffusion approximation even in the linear regime.

There are three ways to do this.
One is to solve the transport equation with appropriate 
collision operators, as was developed by Kirk \& Schneider (1987, 1988)
for relativistic shocks, 
and by Malkov \& V\"{o}lk \shortcite{MV95}
for the injection problem. 
They solved for eigenvalues and eigenfunctions of the transport 
equation both for upstream and downstream 
and by using the matching condition at the shock front 
they obtained the spectral index $\sigma$ and the distribution function. 
This method works well when the pitch angle diffusion is a main 
process of the scattering, while it seems to have met some 
difficulties when the simpler case of pure large angle scattering is 
considered. 

Another approach is the single-particle approach, which was originally 
used by \Bell78 for non-relativistic shocks 
and was extended by \Peacock81 for relativistic shocks. 
Although this method is in principle equivalent to the former method, 
it is more intuitive and more tractable as will be described 
in this paper.
To determine the pitch angle distribution at the shock front, 
\Bell78 and \Peacock81 basically utilized
the distribution functions obtained 
by macroscopic methods such as the diffusion equation.
\Peacock81 also used an upstream distribution function, which is 
realized far upstream when particles suffer large angle scattering;  
these functions may be different from real ones at the shock front 
as was shown in Kirk \& Schneider (1987, 1988). 

The third one is the Monte Carlo simulation in which 
the motion of each particle is traced faithfully
(Ellison, Jones \& Reynolds 1990; Ostrowski 1991;
Bednarz \& Ostrowski 1998).
While Monte Carlo simulation can make a direct estimation of 
the distribution function and the spectral index, a large scale 
simulation is needed to obtain sufficiently accurate results.
Physical interpretations of simulation results 
in terms of analytical models are desirable.

The results of acceleration in such situations
in fact depend on the nature of scattering mechanisms of particles
by magnetic irregularity existing in the background plasma.
This is because
the anisotropy of particle distribution at the shock front
is strongly subject to the detail of the scattering process.
Especially,
two conventional scattering models,
large angle scattering and pitch angle diffusion,
lead to significantly different spectral index
in relativistic or highly oblique shocks
(Kirk \& Schneider 1988; Ellison, Jones \& Reynolds 1990;
Naito \& Takahara 1995; Ellison, Baring \& Jones 1996).
It also may be the case for acceleration of low energy particles.
The pitch angle diffusion is derived by
the quasi-linear theory of plasma turbulence (e.g., Melrose 1980)
but this description is adequate only for weak turbulence
while it is often supposed that the turbulence can be strong
in astrophysical shock environments
(see Ellison, Jones \& Reynolds 1990 and references therein).
On the other hand,
the large angle scattering,
which isotropizes the particle direction in a single scattering
independently of the direction before scattering,
may mimic some effect of strong turbulence,
though there is no theoretical justification for this.
Theory of particle transport in strongly turbulent field
required for realistic investigation of this problem
is very difficult and not yet established,
to our knowledge.

In this paper,
we adopt the large angle scattering model
regarding that it
mimics the scattering under strongly turbulent field,
and investigate the nature of the shock acceleration
in highly anisotropic situation,
basically taking the second approach mentioned above.
The specific large angle scattering model used in the paper is
prescribed in Section2,
and assumes that
the particle direction after scattering does not depend on
the direction before scattering
(although this feature is not always true for general large angle scattering,
we use the term `large angle scattering' for the specific model in this paper).
This mathematical simplicity allows us to treat the problem of particle transport as a random walk.
Thus,
our approach is based on the theory of random walk in a moving medium
not on the diffusion theory
to examine
the pitch angle distribution at the shock front and
the return probability of the particles from the downstream of the shock.
Therefore,
it can be applied to any value of the shock velocity. 
The paper is organized as follows.
In \Refsec{random walk},
we show that
the motion of a particle suffering from 
the large angle scattering can be treated as 
a random walk in a moving medium,
and formulate the random walk using the probability theory.
In \Refsec{approximate solutions}, we derive analytical approximate solutions,
and devote to a specific case of isotropic scattering in \Refsec{results}.
Finally, we apply these results to the shock acceleration  
in \Refsec{shock acceleration}.

\section{Method}
\label{sec:random walk}
In this paper,
we treat the first order Fermi acceleration in parallel shocks
for any value of the shock speed.
We assume the test particle approximation
and adopt large angle scattering 
as the scattering process of particles.

In order to know how particles are accelerated,
the properties of particle trajectories are to be investigated in detail.
When particles move around on each side of the shock front, 
their motion can be regarded as one-dimensional
because, in plane-parallel shocks,
only the position along the field line is relevant.
If the scattering process is assumed to be described by 
large angle scattering, 
the motion of particles on each side of the shock front
can be treated as a random walk
in a moving medium as is described in this section.
Thus, the problem will be reduced to the random walk of particles
in a moving medium with an absorbing boundary; 
the shock front becomes such a boundary both for upstream 
and for downstream. 

In this section, first,
we formulate a random walk treatment
of the motion of particles following the large angle scattering
and examine the properties of the random walk in a moving medium.
Although the method described in this section may also be used for
general problems of particle motion in a moving medium
if this scattering model is suitable,
we confine our attention to the Fermi  
acceleration in shock waves in this paper.

In the following argument, we assume that
the velocity of scattering centres is uniform on each side of 
the shock front and is be equal to the velocity of fluid
(therefore, we call the velocity of the scattering centres
the fluid velocity in the following). 
It is also assumed that energy loss mechanisms of particles 
(e.g. synchrotron loss) is negligible,
and that the scattering of particles is elastic in the fluid frame,
i.e., the particle energy does not change upon scattering.

Hereafter,
we take the unit $c=1$ ($c$ is the speed of light)
and take the spatial coordinate $x$ along
the magnetic field line or the shock normal toward downstream 
direction so that the fluid velocity, $\Vf$,
is positive on each side of the shock front.
We use subscript (+) and (-)
to denote the downstream direction ($+x$)
and upstream direction ($-x$), respectively.
Physical quantities measured in the fluid frame are 
expressed with a prime.

\subsection{Large angle scattering model}
We prescribe the large angle scattering model in the fluid frame
by the following three conditions.
\begin{enumerate}
\item
The energy of particle measured in the fluid frame
is conserved at scattering.
Thus, while particles stay in either the upstream or the downstream 
region, particle speed $v'$ does not change in the fluid frame.
\item
The probability density of displacement of the particle
along the magnetic field line
between successive scatterings is assumed
to obey an exponential distribution
with mean free path $\lambda = \lambda(v', \mu')$.
(Here, the term `the mean free path' means
the displacement along a magnetic field
line and {\it not} a length on the orbit of its gyro motion.)
Then, the probability density of the displacement $\Dx'$
is written as 
\begin{equation}
p(\Dx'; v',\mu') d\Dx' = \left\{
\begin{array}{@{\,}ll}
 \frac{1}{\lambda} e^{-\frac{|\Dx'|}{\lambda}} d\Dx' & 
  \mbox{($\mu'\cdot\Dx' \ge 0$)}\\
 0 & \mbox{(otherwise)},
\end{array}
\right.
\end{equation}
where
the signs of $\Dx'$ and $\mu'$ must be the same.
\item
The pitch angle cosine of particle $\mu'$ after scattering is 
determined according to the probability density function 
$\Pmu(\mu';v')$ independently of the pitch angle before scattering.
Below, for simplicity, we represent this function as $\Pmu$.
Clearly, it must satisfy
$\int_{-1}^{1} \Pmu d\mu' = 1$.
\end{enumerate}
The functional form of $\lambda(v',\mu')$ and $\Pmu(\mu';v')$
still is free and various scattering processes can be adopted.
An often used case is that the mean free time is independent of 
$\mu'$ and the scattering is isotropic,
which we consider in \Refsec{results}.

Because the problems of particle transport
with a boundary are to be considered
in the following,
it is better to describe the position of the particle
in the reference frame
in which the coordinate of the boundary position takes a fixed value,
and scattering centres are flowing with fluid speed $\Vf$.
We call this reference frame the boundary rest frame. 
(For the shock acceleration,
this frame is of course the shock rest frame
for each side of the shock front.
But we consider general case here.)
Therefore, below,
we describe the speed $v'$ and pitch angle cosine $\mu'$ of particles 
in the fluid frame
and describe the position of particle $x$ and the time $t$ in 
the boundary rest frame.
(Since we consider only steady states below,
we need not consider about $t$ in the following in fact.) 
This treatment is widely used for various transport equations
such as diffusion convection equation or radiation hydrodynamics
(Kirk, Schlickeiser \& Schneider 1988).
Thus, we describe the scattering law 
in terms of the displacement measured in the boundary rest frame, 
$\Dx$, which is given through a Lorentz transformation of $\Dx'$ as 
\begin{eqnarray}
\Dx &=& \Gf (\Dx' + \Vf \Dt')
     = \Gf \Dx' (1+\frac{\Vf}{v' \mu'}) \nonumber \\
    &=& \Gf \Dx' (1+\frac{\nu}{\mu'})
\end{eqnarray}
where $\nu$ is defined as
\begin{equation}
\nu := \frac{\Vf}{v'},
\end{equation}
and  $\Gf = 1/\sqrt{1-\Vf^2}$.
Here, $\nu$ represents the degree of advection effects
to the particle motion in the boundary rest frame
and is a fundamental parameter in the random walks in a moving medium
described below.
Clearly, if $\nu \ge 1$,
a particle cannot advance against the flow of scattering centres, 
which is not of interest for the shock acceleration. 
Hence, we consider only the case of $\nu < 1$ below.

It can be shown easily that
when a particle moves towards
(+) direction $\Dx > 0 (\mu>0)$ in the boundary rest frame,
the range of the pitch angle cosine in the fluid frame
is $-\nu < \mu' < 1$,
and for (-) direction $\Dx < 0 (\mu < 0)$,
the range becomes $-1 < \mu' <-\nu$.
Then, the probability distribution of $\Dx$ becomes
\begin{equation}
\label{eq:P(x,mu)}
p(\Dx; v',\mu') d\Dx = \left\{
\begin{array}{@{\,}ll}
 \frac{1}{\lmd} e^{-\frac{|\Dx|}{\lmd}} d\Dx & 
  \mbox{($(\mu'+\nu) \cdot \Dx > 0$)}\\
 0 & \mbox{(otherwise)}
\end{array}
\right.
\end{equation}
where
\begin{equation}
\label{eq:lambda*}
\lmd = \lmd(v', \mu') := \Gf \lambda(v',\mu') |1 + \frac{\nu}{\mu'}|.
\end{equation}
Here, $\lmd$ denotes the mean free path of the particle measured 
in the boundary rest frame
and can depend on $\mu'$ even if $v'$ is the same.
For later argument,
we define the following two scale lengths,
\begin{equation}
  \begin{array}{@{\,}ll}
    \Lp := \max(\lmd; -\nu < \mu' < 1),\\
    \Lm := \max(\lmd; -1 < \mu' < -\nu)
  \end{array}
\end{equation}
where $\Lp$ and $\Lm$ denote
the scale of scattering length in the boundary rest frame
for (+) direction and for (-) direction, respectively.

It should be noted that,
even if both $\lambda$ and $\Pmu$ depend on $v'$,
when the dependences are separable into $v'$ and $\mu'$, 
the properties of particle motion is determined only by $\nu$
except that the temporal and spatial scales can be different.
As far as only the probability of return
and the pitch angle distribution at the shock front are to be considered,
these scales are not relevant at all.
Then, one can choose the scale of the time and space
in order for the computation to become the simplest.

\subsection{Random walk of particles in a moving medium}
In the large angle scattering model
described in the previous subsection,
since the pitch angle of a particle after scattering
is determined independently of that of 
before scattering and $v'$ is conserved,
the displacement of the particle between successive scatterings
is determined independently and equally on each scattering.
Because of this property,
the particle motion can be treated as
a well-defined random walk.
In this subsection, 
we formulate such a random walk treatment. 
Our arguments as to the random walk in this subsection
are based on the Chapter 2 of Cox \& Miller (1965).

\subsubsection{Random walk for large angle scattering models}
In the random walk considered here,
the displacement of the particle $\Dx$
takes a continuous value rather than a discrete value.
For such a random walk, one of the fundamental concepts 
is the probability density function (p.d.f.)
of the displacement for each step of random walk, $f(\Dx)$.
For the large angle scattering model,
we can write this function in terms of $p(\Dx; v',\mu')$ and $\Pmu$,
which introduced in the previous subsection.
Using \Refeq{P(x,mu)}, we obtain
\begin{eqnarray}
\label{eq:pdf}
f(\Dx) &=& \int_{-1}^1 \Pmu p(\Dx; v', \mu') d\mu' \nonumber \\
       &=&
\left\{
\begin{array}{@{\,}ll}
 \displaystyle \int^1_{-\nu} \frac{\Pmu}{\lmd} e^{-\frac{\Dx}{\lmd}} 
d\mu'& (\Dx > 0)\\
 \displaystyle \int^{-\nu}_{-1} \frac{\Pmu}{\lmd} e^{\frac{\Dx}{\lmd}} 
d\mu' & (\Dx < 0) .\\
 \end{array}
\right.
\end{eqnarray}
In this description,
the renewal of $\mu'$ at the scattering
and the translation to the following scattering point
are treated as {\it one step} of the random walk.
Although this p.d.f. have no information about $\mu'$,
if we combine the information of the 
random walk with the scattering law,
the information about $\mu'$ can be taken out as described later.

Corresponding to the p.d.f. $f(\Dx)$,
the moment generating function (m.g.f.) is defined as
\begin{eqnarray}
\label{eq:mgf}
f^*(\theta)
&:=& \int^\infty_{-\infty} e^{-\theta \Dx} f(\Dx) d\Dx \nonumber \\
&=& \int_{-1}^{-\nu} \frac{\Pmu}{1 - \lmd \theta} d\mu'
    + \int_{-\nu}^1 \frac{\Pmu}{1 + \lmd \theta} d\mu',
\end{eqnarray}
which is the two-sided Laplace transform of the p.d.f. $f(\Dx)$.
Clearly, $f^*(0) = 1$.
In general,
at some values of $\theta$ this integral diverges.
For example, for the large angle scattering model,
the value of $\theta = \frac{1}{\Lm}$ and $\theta = - \frac{1}{\Lp}$
are such divergence points.
However, in the following arguments,
we need to consider only the value of $\theta$ in the range,
\begin{equation}
\label{eq:range_th0}
 -\frac{1}{\Lp} < \theta < \frac{1}{\Lm},
\end{equation}
and so we need not worry about such a divergence.

We can derive the following formulas
about the derivatives of $f^*(\theta)$ and the moments of $f(\Dx)$
from m.g.f. as
\begin{eqnarray}
\label{eq:deriv}
\frac{d^n f^*(\theta)}{d \theta^n}
&=& n! \left\{
  \int_{-1}^{-\nu} \frac{\Pmu \lmd^n}{(1 - \lmd \theta)^{n+1}} d\mu'
  \right.
  \nonumber \\
&+& \left. (-1)^n \int_{-\nu}^{1}
    \frac{\Pmu \lmd^n}{(1 + \lmd \theta)^{n+1}} d\mu'
\right\},
\end{eqnarray}
\begin{eqnarray}
\overline{\Dx^n}
&:=& \int_{-\infty}^{\infty} \Dx^n f(\Dx) d\Dx
= (-1)^n \frac{d^n f^*}{d \theta^n}|_{\theta=0} \nonumber \\
&=& n! \left\{ (-1)^n \int_{-1}^{-\nu} \Pmu \lmd^n d\mu'
                  + \int_{-\nu}^{1} \Pmu \lmd^n d\mu'  \right\}.
\nonumber \\ & &
\end{eqnarray}
In the range of $\theta$ mentioned above \refeq{range_th0}, 
$\frac{d^2f^*}{d\theta^2} > 0$ always holds
and then $f^*(\theta)$ is a downward-convex function in this 
region. Therefore, the value of $\theta = \th0 \not = 0$ at 
which $f^*(\th0)=1$ always exists.
This quantity $\th0$,
which is determined only by the functional form of $f(\Dx)$,
includes the most important characteristics 
of the random walk in a moving medium.
In fact,
this $\th0$ together with $\nu$ play an important role
in the following description.
Clearly,
the sign of $\th0$ is equal to the sign of $\nu$ and 
$\overline{\Dx}$, i.e., it is positive throughout this paper.

\subsubsection{Random walk with absorbing barriers}
The problem of random walk in a moving medium with boundaries  
can be formulated as the problem of
{\it the random walk with absorbing barriers}
in the probability theory.
The procedure is as follows.
First, consider that a particle begins a random walk from 
the origin $x=0$ under the situation
in which there are two absorbing barriers at the position 
$x=a$ and $x=-b$ $(a,b>0)$.
Then the particle moves according to the p.d.f. $f(x)$ at each step.
If the particle advances beyond one of the barriers,
it is absorbed by the barrier and the random walk is terminated then.

To make the argument clear,
we introduce the probability density function
of scattering points at the $m$-th step, $f_m(x)$.
The integration of $f_m(x)$ outside of the boundaries ($x<-b$ or $x>a$)
denotes the probability that the particle is absorbed at the $m$-th step.
The integral of $f_m(x)$ over the inside region is {\it not} normalized 
to unity but denotes the probability that the particle is not absorbed 
until or at the $m$-th step and can carry out the $m+1$ step.
The function $f_m(x)$ can be written as a convolution of $f_{m-1}(x)$ 
by $f(x)$ over the non-absorption region,
\begin{equation}
\label{eq:convolution}
f_m(x)
= \int_{-b}^{a} f(x-x') f_{m-1}(x') dx'.
\end{equation}
Because the particle starts from $x=0$,
it can be seen easily
$f_0(x) = \delta(x)$ and $f_1(x) = f(x)$.
All $f_m(x)$ can be calculated in principle 
using \Refeq{convolution} iteratively.

Next,
we introduce the density of scattering points summed over all steps,
\begin{equation}
\label{eq:n(x)}
n(x) := \sum_{m=1}^{\infty} f_m(x) \qquad (-b < x < a),
\end{equation}
which constitutes one of the key concepts in the following description.
If one is not discussing the number of steps before absorption,
the only relevant quantity is the density of scattering points $n(x)$.
We limit our considerations to those using $n(x)$ in this paper, 
although the distribution of the number of steps before absorption is 
important to discuss the dispersion of acceleration time of particles. 
It is noted that $n(x)$ is not the same as the physical number density of particles 
and their mutual relation is explained in the Appendix A.  

From \Refeq{convolution} and \Refeq{n(x)},
we can derive a Fredholm integral equation of the second kind for $n(x)$:
\begin{equation}
\label{eq:n_Integral}
n(x) = f(x) + \int_{-b}^{a} f(x-x') n(x') dx'.
\end{equation}
Especially, for no-boundary case ($a,b \rightarrow \infty$),
the solution $\nN(x)$ obeys
\begin{equation}
\label{eq:nN_Integral}
\nN(x) = f(x) + \int_{-\infty}^{\infty} f(x-x') \nN(x') dx'.
\end{equation}

It is noted that $n(x)$ satisfies certain integral conditions.
To simplify the following description,
we introduce the probability density of absorption positions,
\begin{equation}
\label{eq:tilde_n_org}
\tilde{n}(X)
= f(X) + \int_{-b}^{a} f(X-x) n(x) dx.
\end{equation}
Here $X$ instead of $x$ is used for the spatial coordinate in the 
absorbing regions.
$\tilde{n}(X)$ 
satisfies the following two conditions.
Since the particle is eventually absorbed
by either of the absorbing barriers,
\begin{equation}
\label{eq:PC}
\int_{-\infty}^{-b} \tilde{n}(X) dX + \int_{a}^{\infty} \tilde{n}(X) dX = 1.
\end{equation}
In addition,
in the problem of the random walk with absorbing barriers,
a theorem which is called the Wald's identity holds.
One of the simplest versions of this theorem is the following identity
\begin{equation}
\label{eq:Wald} 
\int_{-\infty}^{-b} \tilde{n}(X) e^{-\th0 X} dX
+ \int_{a}^{\infty} \tilde{n}(X) e^{-\th0 X} dX = 1,
\end{equation}
where $\th0$ was defined in the previous subsection.
In the shock acceleration,
the barrier exists only at one side,
i.e., $a \to \infty$ for downstream region
and $b \to \infty$ for upstream region.
In the following section,
we use one of these conditions
to normalize the approximate solutions.
For $a \rightarrow \infty$,
we use \Refeq{Wald}, i.e.,
\begin{equation}
\label{eq:a_inf}
\int_{-\infty}^{-b} \tilde{n}(X) e^{-\th0 X} dX = 1. 
\end{equation}
On the other hand,
for $b \rightarrow \infty$,
we use \Refeq{PC}, i.e.,
\begin{equation}
\label{eq:b_inf}
\int_{a}^{\infty} \tilde{n}(X) dX = 1.
\end{equation}
%

\subsubsection{No-boundary solution and diffusion length}
In order to know the properties of random walk after large numbers 
of steps, it is helpful to consider the asymptotic form of $\nN(x)$
for $x \rightarrow +\infty$ or $x \rightarrow -\infty$.
Using the convolution theorem to \Refeq{convolution},
the two-sided Laplace transform of $f_m(x)$ for the 
no-boundary case turns out to be 
\begin{equation}
f_m^*(\theta) = \{f^*(\theta)\}^m.
\end{equation}
Therefore, from \Refeq{n(x)},
the two-sided Laplace transform of
the scattering point density for no boundary case $\nN(x)$
within the range $0 < \theta < \th0$ becomes 
\begin{equation}
\nN^*(\theta) = \frac{f^*(\theta)}{1 - f^*(\theta)}.
\end{equation}
Thus, $\nN(x)$ is calculated by the Bromwich integral as
\begin{equation}
\nN(x) = \frac{1}{2 \pi i}
\int_{\sigma -i\infty}^{\sigma +i\infty} \nN^*(\theta) 
e^{\theta x} d\theta,
\end{equation}
where $\sigma$ is real and $0 < \sigma < \th0$.
For $x \rightarrow +\infty$,
the integral path is taken on the $-\theta$ side
and the residue at $\theta = 0$ is dominant in the integral.
The residue is calculated easily as
\[
a_{-1}(\theta = 0) = -\frac{1}{\left. \frac{d f^*(\theta)}{d \theta} 
\right|_{\theta = 0}}.
\]
Then,
we obtain the asymptotic form of $\nN(x)$ for $x \rightarrow +\infty$ as
\begin{equation}
\label{eq:asymptotic_nN_p}
\nN(x)
=- \frac{1}{\left. \frac{d f^*(\theta)}{d \theta} \right|_{\theta = 0}}
= \frac{1}{\overline{x}}
=: n_0
\qquad
\mbox{ for $x \rightarrow +\infty$ }.
\end{equation}
The asymptotic form for $x \rightarrow -\infty$
is calculated in a similar way.
The integral path is taken on the $+\theta$ side
and the residue at $\theta = \th0$ is dominant in the integral.
The residue is calculated as
\[
a_{-1}(\theta = \th0)
=\frac{1}{\left. \frac{d f^*(\theta)}{d \theta} 
\right|_{\theta = \th0}} e^{\th0 x}.
\]
Then, we obtain
\begin{equation}
\label{eq:asymptotic_nN_m}
\nN(x)
= \frac{1}{\left. \frac{d f^*(\theta)}{d \theta} 
\right|_{\theta = \th0}} e^{\th0 x}
=: n_0' e^{\th0 x}
\qquad
\mbox{ for $x \rightarrow -\infty$ }.
\end{equation}
For later convenience,
we define the ratio of these two constants as 
\begin{equation}
\rho := \frac{n_0'}{n_0}
= -\frac{\left. \frac{d f^*(\theta)}{d \theta} \right|_{\theta = 0}}
       {\left. \frac{d f^*(\theta)}{d \theta} \right|_{\theta = \th0}}.
\end{equation}
If $\th0 \Lm, \th0 \Lp \ll 1$ holds,
$\rho$ is almost unity
(see \Refeq{deriv}).
However,
it is to be noted that
$\rho$ is not equal to unity, i.e., $n_0 \neq n_0'$
in general.

Equation \refeq{asymptotic_nN_m} indicates that
the scale length of density of scattering points far upstream
is equal to the inverse of $\th0$.
We denotes this length as
\begin{equation}
\label{eq:LD}
\LD := \frac{1}{\th0}.
\end{equation}
While $\Lp$ and $\Lm$ are the scale lengths of single scattering,
$\LD$ is the scale length of multiple scatterings.
If $\nu$ is small enough and the scattering is isotropic, 
this scale length $\LD$ should coincide
with the diffusion length derived by the conventional diffusion equation, 
$\LDz = \kappa/\Vf$,
where $\kappa$ is the diffusion coefficient 
($\kappa = \frac{1}{3} \lambda v'$), 
which is explicitly shown in \Refsec{results}.
In contrast, when $\nu$ is large,
we can not safely use the conventional diffusion equation
and the length $\LDz$.
However, even in such a case, 
we can regard $\LD$ as a diffusion length,
since the distribution is an outcome of multiple scatterings 
in the random walk. 
Here and hereafter, we call this $\LD$ simply `the diffusion length'
and call $\LDz$ `the conventional diffusion length'.
We also show in \Refsec{results} that
this diffusion length $\LD$ indeed coincides
with the scale length of particle distribution
derived by Peacock (1981) and Kirk \& Schneider (1988) 
in the far upstream region for relativistic shocks.

For later convenience,
we also define the ratio of scale lengths of single scattering to 
the diffusion length as 
\begin{equation}
\Rp := \frac{\Lp}{\LD}
\; , \qquad
\Rm := \frac{\Lm}{\LD}.
\end{equation}
%

\subsubsection[]{Probability density of pitch angle cosine at\\* absorption}
As already mentioned, although
the density of scattering points does not give direct information 
about pitch angle cosine of the particle $\mu'$,
we can derive the probability density of $\mu'$ at absorption 
with an aid of the scattering law. 
The probability density of $\mu'$ at the absorption by the boundary 
$x=a$ is given by 
\begin{equation}
\label{eq:P_a}
P_a(\mu')
= \Pmu e^{-\frac{a}{\lmd}} \left\{ 1 + \int_{-b}^a 
e^{\frac{x}{\lmd}} n(x) dx \right\}.
\end{equation}
and that at the boundary $x=-b$ is given by 
\begin{equation}
\label{eq:P_b}
P_b(\mu')
= \Pmu e^{-\frac{b}{\lmd}}
    \left\{ 1 + \int_{-b}^a e^{-\frac{x}{\lmd}} n(x) dx \right\}.
\end{equation}

The integral conditions for $n(x)$ can be rewritten as
the conditions for $P_a(\mu')$ or $P_b(\mu')$.
For $b \rightarrow \infty$,
the condition \refeq{b_inf} can be rewritten 
in terms of $P_a(\mu')$ as
\begin{equation}
\label{eq:inf_b2}
\int_{-1}^{-\nu} P_a(\mu') d\mu' = 1.
\end{equation}
Similarly,
for $a \rightarrow \infty$,
the Wald's identity \refeq{a_inf} can be rewritten as
\begin{equation}
\label{eq:Wald2}
\int_{-1}^{-\nu} \frac{P_b(\mu')}{1 - \th0\lmd} d\mu' = e^{-\th0 b}.
\end{equation}
Here,
we define the total absorption probability by the boundary at $-b$
\begin{equation}
\Pabs(b) := \int_{-1}^{-\nu} P_b(\mu') d\mu'.
\end{equation}
This function also means the probability that the particle,
which is injected in downstream side at a distance of $b$ from the 
boundary, reaches the boundary against the flow of scattering centres.
If $\Rm \ll 1$ holds, \Refeq{Wald2} also means
\begin{equation}
\label{eq:Pabs_D}
\Pabs(b) = e^{-\th0 b}
\end{equation}
for {\it any} models.
Since, for isotropic scattering and for $\nu \rightarrow 0$,
$\LD = 1/\th0$ approaches the conventional diffusion length $\LDz$
as already mentioned,
\Refeq{Pabs_D} agrees with
the result of the conventional diffusion approximation
obtained by Drury (1983).

\subsection{The return probability}
In order to apply the random walk theory described above to the shock acceleration,
it is useful to consider the probability density of the 
pitch angle cosine $\mu'$ when the particle which crossed the shock 
front with $\mu_0'$ returns to the front crossing the reverse direction.
For downstream, random walk with a finite $b$ and $a \to \infty$ 
is applied while for upstream that with a finite $a$ and $b \to \infty$ 
is applied. 
As will be shown in later sections,
all the quantities associated with the shock acceleration
can be represented with this probability density.
Hereafter, we denote the spatial coordinate by $z$ with the shock front 
at $z=0$ in order to retain $x$ coordinate for the random walk 
problem treated so far where $x=0$ corresponds to the 
injection point of the particle. 

First, let us consider the return probability for the upstream. 
The particle first crosses the boundary toward (-) direction, i.e., 
from the downstream to the upstream
with a pitch angle cosine $\mu'_0 (< -\nu)$.
In order to calculate the probability density of the pitch angle 
cosine at the return $\mu' (> -\nu)$, $\Pm(\mu'; \mu'_0)$,
it is useful to introduce the density of scattering points   
before return, $\Nm(z;\mu'_0)$, which is defined on $z<0$.
This density can be calculated in the following way.
The probability density of the first scattering point for 
the particle which crossed the boundary with pitch angle cosine $\mu'_0$
is determined by the scattering law \refeq{P(x,mu)}.
Particles which have the first scattering point at $z=-a$
make the scattering point density $n(z+a;a,\infty)$, 
where $n(x;a,b)$ denotes the scattering point density of 
particles injected at $x=0$ in the random walk with absorbing 
barriers at $x=a$ and $x=-b$. 
Therefore, we obtain
\begin{equation}
\label{eq:Nm}
\Nm(z;\mu'_0)
=  \frac{1}{\lmdz} e^{\frac{z}{\lmdz}}
 + \frac{1}{\lmdz} \int_{0}^{\infty} n(z+a; a, \infty) 
e^{-\frac{a}{\lmdz}} da
\end{equation}
where $z < 0$ and $\lmdz = \lmd(\mu'_0)$.
Using this,
$\Pm(\mu'; \mu'_0)$ is calculated as
\begin{equation}
\label{eq:Pm}
\Pm(\mu'; \mu'_0)
= \Pmu \int_{-\infty}^{0} \Nm(z; \mu'_0) e^{\frac{z}{\lmd}} dz
\end{equation}
where $\lmd = \lmd(\mu')$.
For this case,
the total return probability is always unity (\Refeq{inf_b2}), i.e.,
\begin{equation}
\label{eq:total_Pm}
\int_{-\nu}^{1} \Pm(\mu'; \mu'_0) d\mu' = 1. 
\end{equation}
We can rewrite this as an integral condition for $\Nm(z;\mu'_0)$,
\begin{equation}
\label{eq:Nm_condition}
\int_{-\nu}^{1} \Pmu
\int_{-\infty}^{0} \Nm(z;\mu'_0) e^{\frac{z}{\lmd}} dz d\mu'
= 1.
\end{equation}

For the downstream, the probability density of pitch angle cosine 
$\mu' (< -\nu)$ at return for the particle which initially crossed 
the shock front with $\mu'_0 (> -\nu)$ ((+) direction),
$\Pp(\mu'; \mu_0')$, is calculated in a similar way.
The scattering point density $\Np(z; \mu'_0)$ is written as
\begin{equation}
\label{eq:Np}
\Np(z;\mu'_0)
=  \frac{1}{\lmdz} e^{-\frac{z}{\lmdz}}
 + \frac{1}{\lmdz} \int_{0}^{\infty} n(z-b; \infty,b) 
e^{-\frac{b}{\lmdz}} db,
\end{equation}
which is defined on $z > 0$.
Using this density,
$\Pp(\mu'; \mu'_0)$ is calculated as
\begin{equation}
\label{eq:Pp}
\Pp(\mu'; \mu'_0)
= \Pmu \int_{0}^{\infty} \Np(z; \mu'_0) e^{-\frac{z}{\lmd}} dz.
\end{equation}
The total return probability $\Pret(\mu'_0)$ is not unity
and is given as
\begin{equation}
\label{eq:Pret}
\Pret(\mu'_0) = \int_{-1}^{-\nu} \Pp(\mu'; \mu'_0) d\mu'.
\end{equation}
The integral condition \refeq{Wald2} becomes
a condition for $\Np(z;\mu'_0)$ as
\begin{equation}
\label{eq:Np_condition}
\int_{-1}^{-\nu} \frac{\Pmu}{1 - \th0 \lmd}
\int_{0}^{\infty} \Np(z;\mu'_0) e^{-\frac{z}{\lmd}} dz d\mu'
= \frac{1}{1 + \th0 \lmdz},
\end{equation}
as is shown in \Refapn{condition}.

In the following sections,
we derive approximate analytic solutions 
of $\Pm(\mu';\mu'_0)$ and $\Pp(\mu';\mu'_0)$
where we use the normalization condition
for the scattering point density $\Nm$ and $\Np$.

\section{Approximate solutions}
\label{sec:approximate solutions}
In this section, we derive approximate analytic solutions for 
the probability densities by obtaining approximate solutions for 
the integral equation \refeq{n_Integral}.
We start from the asymptotic solution of \refeq{asymptotic_nN_p} 
and \refeq{asymptotic_nN_m} for the no-boundary problem. 
First we approximate the no-boundary solution $\nN(x)$ as
\begin{equation}
\label{eq:approx_nN}
\nN(x) \sim \nNz(x) :=
\left\{
\begin{array}{@{\,}ll}
 \displaystyle n_0' e^{\th0 x} & (x < 0)\\
 \displaystyle n_0 & (x > 0).\\
 \end{array}
\right.
\end{equation}
Note that this approximation neglects a peak around $x=0$ 
arising from the injection (see \Reffig{nNx} in \Refsec{results}
for Monte Carlo results).

To obtain approximate solutions for problems with finite boundaries 
from those for no-boundary problems, we transform 
the original integral equation \refeq{n_Integral} to an alternative 
integral equation
\begin{equation}
\label{eq:n_Integral2}
n(x) = \nN(x) - k(x,0) - \int_{-b}^{a} k(x,x')n(x') dx',
\end{equation}
where we define the kernel
\begin{eqnarray}
\label{eq:kernel}
k(x,x')
&:=& \int_{a}^{\infty} \nN(x-X)f(X-x')dX  \nonumber \\
&+&      \int_{-\infty}^{-b} \nN(x-X)f(X-x')dX
\end{eqnarray}
(see \Refapn{integral_equation} for its derivation).
This integral equation is, of course, equivalent to \Refeq{n_Integral}.

\subsection[]{Approximate solution of the probability\\* densities}
First, we consider approximate solution for the upstream $\Nm(z)$ 
taking $b \rightarrow \infty$ and a finite value of $a$. 
Starting from the approximation \refeq{approx_nN},
the kernel of the integral equation \refeq{kernel} for $n(x)$ 
becomes separable into $x$, $x'$ as
\[
k(x,x') = n_0' e^{\th0 (x-a)} \phi_{-}(x'-a)
\]
where
\[
\phi_{-}(x') := \int_{0}^{\infty} f(X-x') e^{-\th0 X} dX.
\]
Substituting this into \Refeq{n_Integral2},
we obtain
\begin{eqnarray*}
\label{eq:approx_nx_m}
n(x) &=& \nN(x) 
     - n_0' e^{\th0 (x-a)}
     \left\{ \phi_{-}(-a) + C_a \right\} \\
& = & \nN(x)  -n_0'  e^{\th0 (x-a)}C_a'
\end{eqnarray*}
where
\[
C_a := \int_{-\infty}^{a} \phi_{-}(x'-a) n(x') dx'
\]
and
\[
C_a' := \phi_{-}(-a) + C_a.
\]
The term proportional to $C_a'$ represents the effects of 
the existence of the absorbing boundary at $x=a$. 
Although we do not need to explicitly evaluate $C_a$ 
in our procedures described below, 
$C_a$ can be solved if we substitute \refeq{approx_nx_m} into the 
definition of $C_a$ as 
\[
C_a = \frac{1}{1 + n_0' C_2}
\left\{
  \int_{-\infty}^{a} \phi_{-}(x-a) \nN(x) dx
  - n_0' C_2 \phi_{-}(-a)
\right\}
\]
where
\[
C_2 := \int_{-\infty}^{0} \phi_{-}(x') e^{\th0 x'} dx'.
\]

If we use the approximation \refeq{approx_nN} again 
and remembering $\LD=1/\th0$,
the \Refeq{Nm} becomes
\begin{equation}
\label{eq:approx_Nm}
\Nm(z;\mu'_0)
= C_0(\lmdz) \frac{1}{\lmdz} e^{\frac{z}{\lmdz}}
+ C_1(\lmdz) \frac{1}{\LD} e^{\frac{z}{\LD}}
\end{equation}
where
\begin{equation}
\label{eq:C_0}
C_0(\lmdz) := 1 + n_0 \lmdz - \frac{n_0' \lmdz}{1 - \th0 \lmdz},
\end{equation}
and 
\[
C_1(\lmdz) := \frac{n_0'/\th0}{1 - \th0 \lmdz}
- \frac{n_0'}{\th0 \lmdz} \int_{0}^{\infty} C_a'(a) e^{-\frac{a}{\lmdz}} da.
\]
%
In \Refeq{approx_Nm},
the second term with the scale length $\LD$  represents
the distribution formed by the diffusion process
after a large number of scattering,
similar to
the asymptotic form of $\nN(x)$ for far upstream \refeq{asymptotic_nN_m}.
The first term represents
the correction to the diffusive term
by the following three non-diffusive effects; 
the density of scattering points
contributed by particles which take only a few steps,
the effect of finite initial mean free path $\lmdz$
(finite distance from the boundary to the first scattering point),
and
the absorbing effect near the boundary.
When we take the limit $\lmdz \to 0$,
this term disappears and
\Refeq{approx_Nm} is just reduced to the diffusive solution.
Depending on $\lmdz$,
$C_0$ can take a positive or negative value.
For small $\lmdz$
the correction term is positive
and it mainly reflect the scattering points of the few step particles,
while for large $\lmdz$
it is negative
accounting for the effects of finite mean free path
and absorption at the boundary.
These features are mentioned again in \Refsec{results}
for a specific scattering model.

Substituting this to \Refeq{Pm},
we obtain an approximate solution for $\Pm$
\begin{equation}
\label{eq:approx_Pm}
\Pm(\mu'; \mu'_0)
= C_0(\lmdz) \frac{\Pmu \lmd}{\lmd + \lmdz}
+ C_1(\lmdz) \frac{\Pmu \lmd}{\lmd + \LD}.
\end{equation}
However, this approximate solution does not satisfy
the condition \refeq{total_Pm} in general 
because we do not use the exact solution of $n(x)$. 
Although there may be various ways of remedying this problem, 
here we adopt a simple and practical way to renormalize the 
amplitude of $C_1$ so as to satisfy \refeq{Nm_condition} for 
a given $\lmdz$.
Thus, instead of $C_1$ we use the coefficient $\tilde{C}_1$ 
defined by 
\begin{equation}
\label{eq:C_1}
\tilde{C}_1(\lmdz) = \frac{1 - C_0 \gp(\lmdz)}{\gp(\LD)}
\end{equation}
where we define
\begin{equation}
\label{eq:gp}
\gp(l) := \int_{-\nu}^{1} \frac{\Pmu \lmd}{\lmd + l} d\mu'.
\end{equation}
This method also has a merit in avoiding to calculate 
a complicated integral in the original definition of $C_1$.

Next, we consider approximate solution for the downstream $\Np(z)$
in a similar way by taking $a \rightarrow \infty$ 
and a finite value of $b$. 
Starting from the approximation \refeq{approx_nN}, 
the kernel becomes 
\[
k(x,x') = n_0 \phi_{+}(x')
\]
where
\[
\phi_{+}(x) := \int_{-\infty}^{-b} f(X-x) dX.
\]
$n(x)$ is obtained as 
\begin{equation}
\label{eq:approx_nx_p}
n(x) = \nN(x) - n_0 C_b'
\end{equation}
where
\[
C_b' := \phi_{+}(0) + \int_{-b}^{\infty} \phi_{+}(x') n(x') dx'.
\]
Substituting this into \Refeq{Np}
and using the approximation \refeq{approx_nN} again,
we obtain 
\begin{equation}
\label{eq:approx_Np}
\Np(z;\mu'_0)
= C_0' \frac{1}{\lmdz} e^{-\frac{z}{\lmdz}} + C_1'
\end{equation}
where
\begin{equation}
\label{eq:C_0'}
C_0'(\lmdz) := 1 - n_0 \lmdz + \frac{n_0' \lmdz}{1 + \th0 \lmdz},
\end{equation}
\[
C_1'(\lmdz) := n_0 \left\{ 1 - \int_{0}^{\infty}
               C_b'(b) e^{-\frac{b}{\lmdz}} db \right\}.
\]
Similarly to the case of $\Nm$,
the second term in \refeq{approx_Np} denotes
the diffusive term,
which corresponds to the asymptotic form for far downstream \refeq{asymptotic_nN_p},
and the first term represents the correction to it.
$C_0'$ takes positive value for small $\lmdz$
and negative value for large $\lmdz$.
Using this to \Refeq{Pp},
we obtain approximate solution of $\Pp$
\begin{equation}
\label{eq:approx_Pp}
\Pp(\mu'; \mu'_0)
= C_0'(\lmdz) \frac{\Pmu \lmd}{\lmd + \lmdz} + C_1'(\lmdz) \Pmu \lmd.
\end{equation}
We again renormalize $C_1'$ so as to satisfy the 
integral condition \refeq{Np_condition} for the same reason as 
in the upstream. Thus, instead of $C_1'$ we use $\tilde{C}_1'$ 
defined by 
\begin{equation}
\label{eq:C_1'}
\tilde{C}_1'(\lmdz) = \frac{1 - C_0'(\lmdz) (\gm(\lmdz) + \gp(\LD) )}
{\gp(\LD) (\lmdz + \LD)}
\end{equation}
where we define
\begin{equation}
\label{eq:gm}
\gm(l) := \int_{-1}^{-\nu} \frac{\Pmu \lmd}{\lmd + l} d\mu'
\end{equation}
and we use the relation
\begin{equation}
\label{eq:gp_gm}
-\gm(-\LD) = \gp(\LD),
\end{equation}
which is derived by \Refeq{mgf}
together with the definition of $\th0$
and $\int_{-1}^{1} \Pmu d\mu' = 1$.
The total return probability $\Pret$ is calculated by 
\Refeq{approx_Pp} easily,
\begin{equation}
\label{eq:approx_Pret}
\Pret(\mu'_0) = C_0'(\lmdz) \gm(\lmdz) + \tilde{C}_1'(\lmdz) \hm
\end{equation}
where we define
\begin{equation}
\hm := \int_{-1}^{-\nu} \Pmu \lmd d\mu'.
\end{equation}

\subsection{Multi-step approximation}
To compare our approximation with more conventional diffusive
approximation,
for reference we consider also the latter approximation which we 
call multi-step approximation. 
This is done by ignoring the first term 
in \Refeq{approx_Nm} or \refeq{approx_Np},
i.e., by setting $C_0, C_0' \rightarrow 0$. 
In this approximation,
the density profile takes the same form
to the results of the diffusive approximation.
But,
since we do not take the limit $\lmdz \to 0$,
this approximation includes partly
the effect of the finite mean free path.
(Although, the effect of the first step return is lost.)

Thus, in the multi-step approximation,
the probability density functions become
\begin{equation}
\label{eq:approx_Pm_m}
\Pm(\mu'; \mu'_0)
= \frac{1}{\gp(\LD)} \frac{\Pmu \lmd}{\lmd + \LD},
\end{equation}
\begin{equation}
\label{eq:approx_Pp_m}
\Pp(\mu'; \mu'_0)
= \frac{1}{\gp(\LD)} \frac{\Pmu \lmd}{\lmdz + \LD}
\end{equation}
and
\begin{equation}
\label{eq:approx_Pret_m}
\Pret(\mu'_0) = A \frac{\LD}{\lmdz + \LD}, 
\end{equation}
where we define
\begin{equation}
\label{eq:const_A}
A := \frac{\hm}{\LD \gp(\LD)}.
\end{equation}
In the multi-step approximation, the pitch angle 
distribution at return becomes independent of $n_0$ and $n_0'$.
Further $\Pm$ is independent of $\mu_0'$
and $\Pp$ is separable with respect to $\mu_0'$ and $\mu'$.
Although in general the multi-step approximation is not 
satisfactory, there are cases where 
this makes some sense as shown in \Refsec{results}.

\subsection{The absorption probability}
The absorption probabilities of the particle injected 
at a distance $b$ in the downstream from the shock front,  
$P_b(\mu')$ and $\Pabs(b)$, are of some theoretical interest, 
although it is not explicitly used in later considerations. 
To obtain their approximate expression,
we start with \Refeq{approx_nx_p}
together with the approximation \refeq{approx_nN}.
We renormalize the amplitude of $c:=-n_0 C_b'$
so as to satisfy the Wald's identity \refeq{Wald2} as
\begin{eqnarray}
\tilde{c} &=& \frac{1}{\gp(\LD)}
   \left\{ \frac{1}{\LD} e^{-\frac{b}{\LD}}
   - \int_{-1}^{-\nu} \frac{\Pmu C_0(\lmd) e^{-\frac{b}{\lmd}}}
{\LD - \lmd} d\mu' \right. \nonumber \\
   && - \left. n_0' \LD e^{-\frac{b}{\LD}} \int_{-1}^{-\nu} 
\frac{\Pmu \lmd}{(\LD - \lmd)^2} d\mu'
   \right\}.
\end{eqnarray}
Using $\tilde{c}$, we obtain
\begin{equation}
P_b(\mu')
= \Pmu \left\{ e^{-\frac{b}{\lmd}} C_0(\lmd) 
+ n_0' e^{-\frac{b}{\LD}} \frac{\lmd \LD}{\LD - \lmd}
  + \tilde{c} \lmd \right\},
\end{equation}
\begin{eqnarray}
\Pabs(b)
&=& \int_{-1}^{-\nu} \Pmu C_0(\lmd) e^{-\frac{b}{\lmd}} d\mu' 
\nonumber \\
&&+ \LD \gp(\LD) \left\{ n_0' e^{-\frac{b}{\LD}}  
+ \tilde{c} A \right\}.
\end{eqnarray}
It is to be noted that $\Pabs(b)$ has 
the scale length $\LD$ for large $b$.

The absorption probability of the particle injected 
in the upstream at the distance of $a$ from the shock front 
and the pitch angle cosine $\mu'$,  
$P_a(\mu')$, can be also calculated in a similar manner. 
Using \Refeq{approx_nx_m} with \refeq{approx_nN} we 
renormalize $c' := n_0' C'(a) e^{-\th0 a}$ so as to 
satisfy \Refeq{inf_b2},
although
we do not present the result here.
In principle it is possible to calculate $\Pm$ and $\Pp$ by 
using these $P_b(\mu'), P_a(\mu')$. 
However, since the expressions are too cumbersome and 
impractical to denote $\Pm$ and $\Pp$ in an analytical form,
we use the \Refeq{approx_Pm} and \refeq{approx_Pp}
for $\Pm$ and $\Pp$ respectively
in the following argument.

We make further approximation
such as the multi-step approximation.
If we set $n_N(x) = n_0$ for all $x$,
which is the diffusive solution,
and if we ignore the first term in \Refeq{P_b},
using Wald's identity \refeq{Wald2},
we obtain 
\begin{equation}
P_b(\mu') = A e^{-\frac{b}{\LD}} \frac{\Pmu \lmd}{\hm},
\end{equation}
\begin{equation}
\label{eq:Pabs_m}
\Pabs(b) = A e^{-\frac{b}{\LD}}.
\end{equation}
In the limit of $\nu \rightarrow 0$,
in which the conventional diffusion approximation holds,
the constant $A$ approaches unity
and this expression of $\Pabs(b)$ agrees with \Refeq{Pabs_D}.
Peacock(1981) derived the absorption probability in a similar form
(see his equation (24);
he did not give a concrete 
expression of the constant $A$),
but did it in the downstream fluid frame
under non-relativistic fluid speed condition
and the diffusion approximation.
In the following section,
we will make comparisons with his expression
for the isotropic scattering.

Corresponding approximation in the upstream can be made in a 
similar way. 
If we set the diffusive solution $n_N(x) \propto e^{\frac{x}{\LD}}$, 
$n(x)$ becomes $c' e^{\frac{x}{\LD}}$, where $c'$ is some constant.
Ignoring the first term in \Refeq{P_a} and 
determining $c'$ through \Refeq{inf_b2},
we obtain 
\begin{equation}
P_a(\mu') = \frac{1}{\gp(\LD)} \frac{\Pmu \lmd}{\lmd + \LD}.
\end{equation}
If we use these expressions for $P_b(\mu')$ and $P_a(\mu')$ 
to calculate $\Pp$ and $\Pm$,
the results coincide with
the results of the multi-step approximation
\refeq{approx_Pm_m}, \refeq{approx_Pp_m} obtained in 
the previous subsection as should be since only diffusive 
return is taken into account.
Therefore, we call these expressions for $P_b(\mu')$, $\Pabs(b)$ and 
$P_a(\mu')$  `the multi-step approximation', too.

\section[]{Results for Isotropic Large Angle\\* Scattering Model}
\label{sec:results}
In this section, we apply the random walk theory
developed so far to a specific model of scattering.
We consider here the model in which
the mean free time $\tau$ is independent of $\mu'$
(although the energy dependence may be allowed, it is not 
relevant here), i.e., $\tau(\mu',v') = \tau_0(v')$,
and the scattering is isotropic in the fluid frame,
\begin{equation}
\lambda = v' |\mu'| \tau_0
\; , \qquad
\Pmu = \frac{1}{2}.
\end{equation}
This model has been widely used 
both in analytic works (Peacock 1981) 
and in the Monte Carlo simulations (Ellison, Jones \& Reynolds 1990).  
To check the validity of our model,
we also perform the Monte Carlo simulation
in which the particle position is traced step by step faithfully
according to the adopted scattering law.
In the simulation,
the escaping boundary is set at $X_{\rm esc} = 15 \LD$,
which is far enough not to influence the return probability of 
the particle (see e.g. \Refeq{Pabs_m}).

\subsection[]{Properties of the random walk with isotropic\\* scattering}

%
\begin{figure*}
  \leavevmode
    \epsfysize=8cm
    \epsfbox{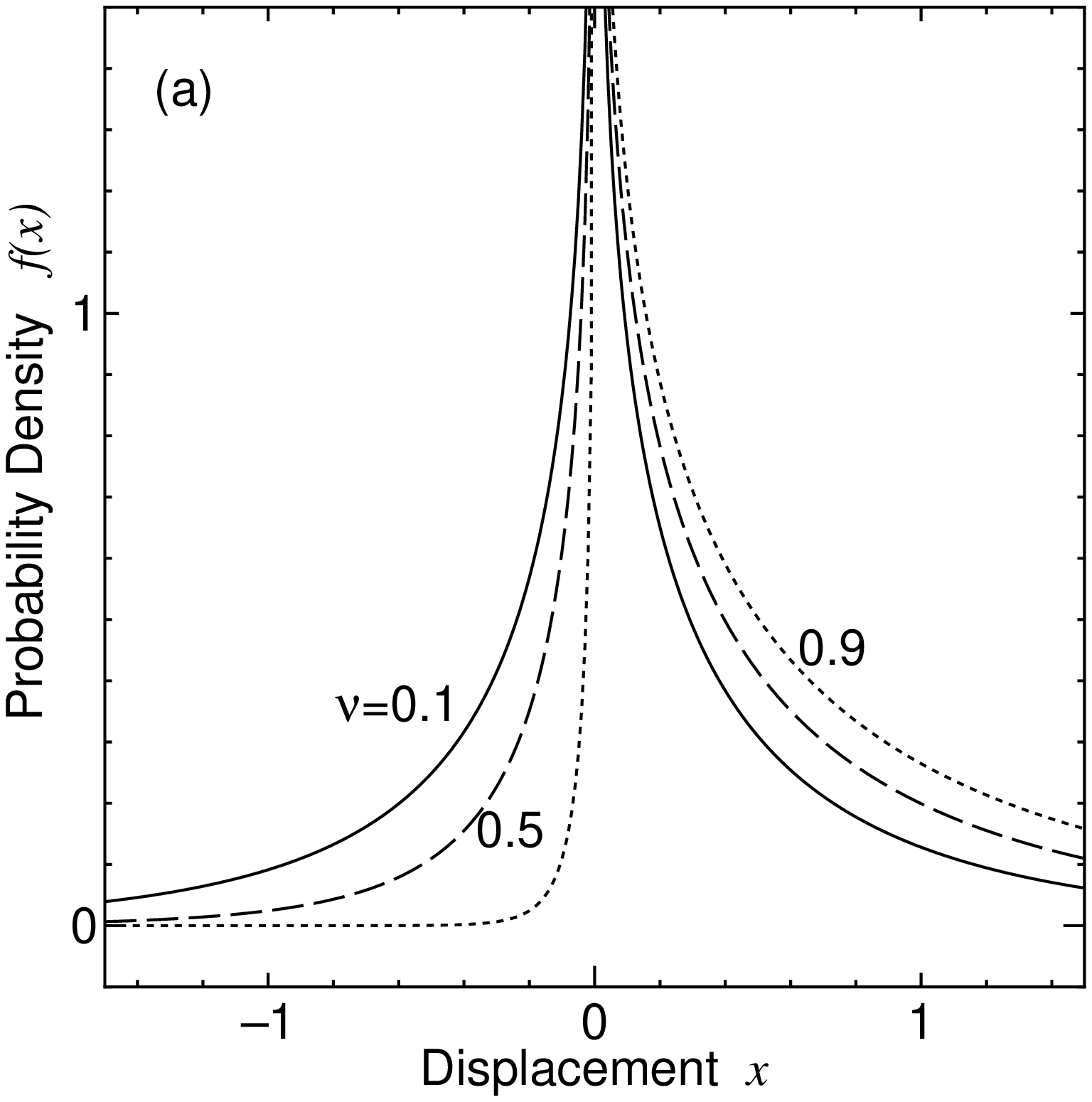}
    \epsfysize=8cm
    \epsfbox{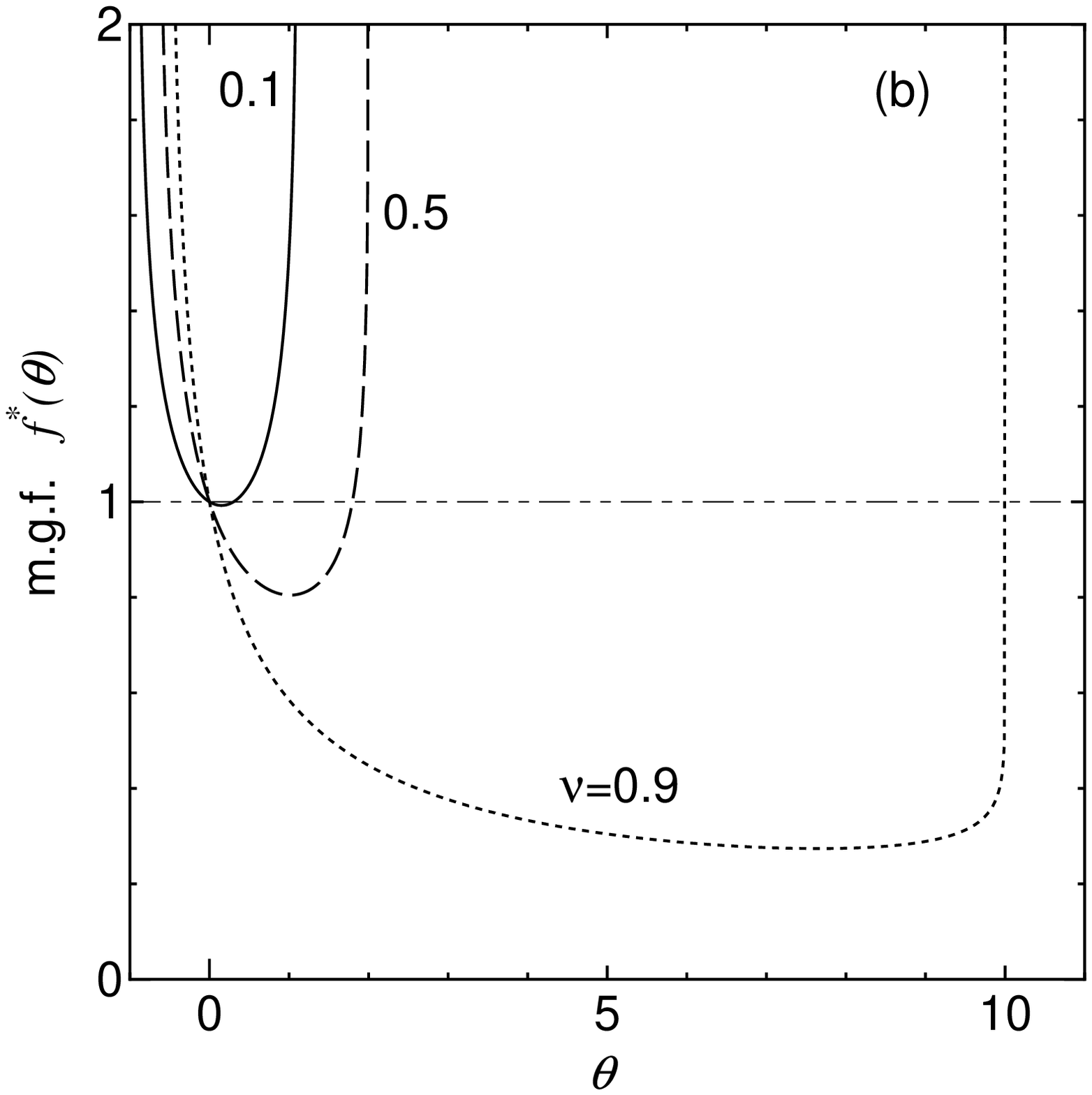}
  \caption{
    (a) The p.d.f. $f(x)$ and (b) the m.g.f. $f^*(\theta)$
    for isotropic large angle scattering model.
    Those for $\nu=0.1$ (solid curve), $0.5$ (dashed curve)
    and $0.9$ (dotted curve) are shown.
  }
  \label{fig:pdf_mgf}
\end{figure*}
In this model, $\lmd$ is given by 
\begin{equation}
\lmd = \Gf v'\tau_0 |\mu' + \nu|.
\end{equation}
For the reason mentioned in section 2.1,
we can take the unit of length as $\Gf v'\tau_0$
in the following argument.
Thus, $\lmd$ becomes
\begin{equation}
\lmd = |\mu' + \nu|
\end{equation}
and
\begin{equation}
\Lp = 1 + \nu \; , \qquad \Lm = 1 - \nu.
\end{equation}
The p.d.f. of the random walk is given by 
\begin{equation}
f(x) =
\left\{
\begin{array}{@{\,}ll}
 \displaystyle \frac{1}{2} E_1(\frac{x}{\Lp}) & (x > 0)\\
 \displaystyle \frac{1}{2} E_1(\frac{|x|}{\Lm}) & (x < 0)\\
 \end{array}
\right.
\end{equation}
where
$E_n(x)$ is the exponential integral defined by
(see Abramowitz \& Stegun \shortcite{AS})
\begin{equation}
E_n(x) := \int_{1}^{\infty} \frac{e^{-xt}}{t^n} dt.
\end{equation}
The m.g.f. is given by 
\begin{equation}
\label{eq:f^*1}
f^*(\theta) = \frac{1}{2 \theta}
\ln\left(\frac{1 + \theta(1+\nu)}{1 - \theta(1-\nu)}\right)
\end{equation}
and
the moments of $f(x)$ are calculated as 
\begin{equation}
\overline{x} = \nu,
\qquad
\overline{x^2} = \frac{1}{3} +\nu^2,
\qquad
\overline{x^2} - \overline{x}^2 = \frac{1}{3}.
\end{equation}
\Reffig{pdf_mgf}
shows the p.d.f. $f(x)$ (a), and the m.g.f. $f^*(\theta)$ (b) 
for $\nu = 0.1, 0.5, 0.9$.
It is seen that the p.d.f. becomes more asymmetric for larger $\nu$
because of the advection effect.

The equation to determine $\th0$ (or $\LD:= 1/\th0$) is given by 
\begin{equation}
\label{eq:th0_1}
2 \th0 - \ln\left(\frac{1 + \th0(1+\nu)}{1 - \th0(1-\nu)}\right) = 0.
\end{equation}
\begin{figure}
  \leavevmode
    \epsfysize=8cm
    \epsfbox{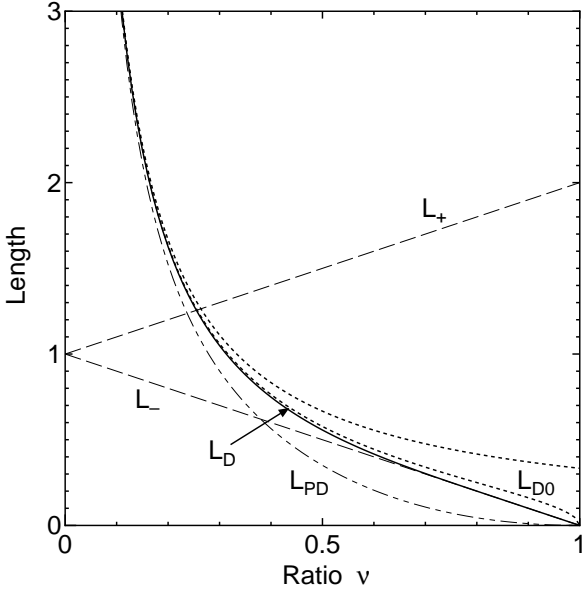}
  \caption{
    The scale lengths of the random walk
    for the isotropic large angle scattering model
    in the unit $\Gf v'\tau_0$.
    The diffusion length $\LD=1/\th0$ (solid curve),
    the scattering length to the upstream direction 
    $\Lm$ (lower dashed curve)
    and that for downstream direction $\Lp$ (upper dashed curve)
    are shown as functions of $\nu$, the ratio of the fluid 
    velocity to the particle velocity.
    The conventional diffusion length $1/(3\Gf\nu)$ ($\nu=\Vf$, lower dotted curve)
    and $1/(3\nu)$ ($\Gf=1$, upper dotted curve) are also shown (see text).
    $L_{\rm PD}$ (dot-dashed curve)
    is the diffusive scale length for
    a model of pitch angle diffusion
    derived by Kirk \& Schneider (1987a).
  }
  \label{fig:scale}
\end{figure}
\Reffig{scale}
presents the scale lengths of the random walk,
$\Lp$, $\Lm$ and $\LD$.
We also show the conventional diffusion length $\LDz$ mentioned in Section 2.2.3
extrapolating it to high $\nu$.
Since
it depends on the Lorentz factor of fluid speed $\Gf$
in the unit of length used here ($\LDz = 1/(3\Gf\nu)$),
we show the two cases;
the case of a relativistic particle in a relativistic flow
($v' \sim 1, \nu \sim \Vf$; lower dotted curve),
and
the case of a non-relativistic particle in a non-relativistic flow
($v',\Vf \ll 1, \Gf \sim 1$; upper dotted curve, $\LDz = 1/(3\nu)$).
The diffusive scale length for a pitch angle diffusion model 
$L_{\rm PD}$ (Kirk \& Schneider 1987a) is represented, too.
(Here,
we choose the pitch angle diffusion coefficient in their model
so that it has the same spatial diffsion coefficient
as our model.)
For $\nu \ll 1$,
$\LD$ can be approximated as
\begin{equation}
\label{eq:th0_1_3V}
\LD \sim \frac{1}{3 \nu},
\qquad
(\th0 \sim 3 \nu)
\end{equation}
and for $\nu \rightarrow 1$,
\begin{equation}
\LD \sim \Lm (1 + \epsilon)
\qquad
(\th0 \sim \frac{1-\epsilon}{1 - \nu})
\end{equation}
where
\begin{equation}
\epsilon := \frac{2}{\Lm} e^{-\frac{2}{\Lm}}.
\end{equation}
Therefore,
if the ratio $\nu$ is small enough ($\nu \la 0.3$),
our diffusion length $\LD$ coincides with the conventional one 
$\LDz$, as expected.
On the other hand,
if $\nu$ becomes larger,
the difference between $\LD$ and $\LDz$ becomes larger.
As mentioned in Section 2.2.3,
for large $\nu$,
$\LDz$ becomes meaningless
because the conventional diffusion approximation is no longer valid.
However,
$\LD$ still has a meaning of `the diffusion length'
and this is explicitly shown later in \Reffig{Pabs}.
In fact,
this scale length has been derived
by \Peacock81 and Kirk \& Schneider (1988) independently
as the scale length of
the particle distribution
on the far upstream region of relativistic parallel shocks
for large angle scattering models.
In \Peacock81,
the equation that is equivalent to \Refeq{th0_1}
was derived from the conservation law of particles in the upstream 
of the shock using `the relativistic diffusion approximation'
(which was named by \KS87),
where $(a + u_1)/(1 + u_1^2)$ corresponds to our $\th0$
in his equation (16).
In Kirk \& Schneider (1988),
such equivalent equation to \refeq{th0_1} was also derived
as the equation to determine the eigenvalues
for large-angle scattering operator.
In their equation (34),
$1/\nu_i$ corresponds to $\th0$
and $\omega$ was taken to unity.
The upstream density profile
obtained in a previous Monte Carlo work
for oblique shocks (Naito \& Takahara 1995)
deviates from the diffusion approximation,
which can be interpreted using $\LD$ obtained here
if we consider in the de Hoffmann-Teller frame.
\Reffig{scale} 
also shows that, when $\nu$ becomes larger,
the difference between $\Lp$ and $\Lm$ is larger
and $\LD$ approaches to $\Lm$.
Behaviour of $L_{\rm PD}$ is similar to that of $\LD$
but takes a smaller value.
It indicates
that the diffusive scale length
depends on the scattering model explicitly.

\begin{figure}
  \leavevmode
    \epsfysize=8cm
    \epsfbox{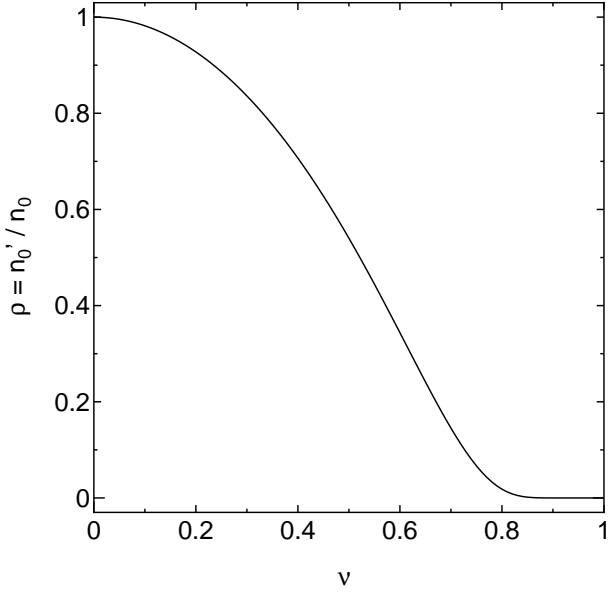}
  \caption{
    The ratio $n_0'$ to $n_0$ as a function of $\nu$.
  }
  \label{fig:rho}
\end{figure}

The constants $n_0$ and $n_0'$, which determine the 
density of scattering points in the no-boundary case,
becomes
\begin{equation}
n_0 = \frac{1}{\nu},
\qquad
n_0' = \frac{1}{\th0(1-\nu^2) - 2\nu} - \th0.
\end{equation}
\Reffig{rho} shows the ratio $\rho = n_0'/n_0$
as a function of $\nu$.
For the limit $\nu \rightarrow 0$,
$\rho$ approaches to unity,
which agrees with the results of the conventional 
diffusion approximation (Drury 1983).
However, as $\nu$ becomes larger,
$\rho$ becomes smaller
and approaches $0$
as $2 \epsilon / \Lm^2$
when $\nu \rightarrow 1$.

\begin{figure}
  \leavevmode
    \epsfysize=8cm
    \epsfbox{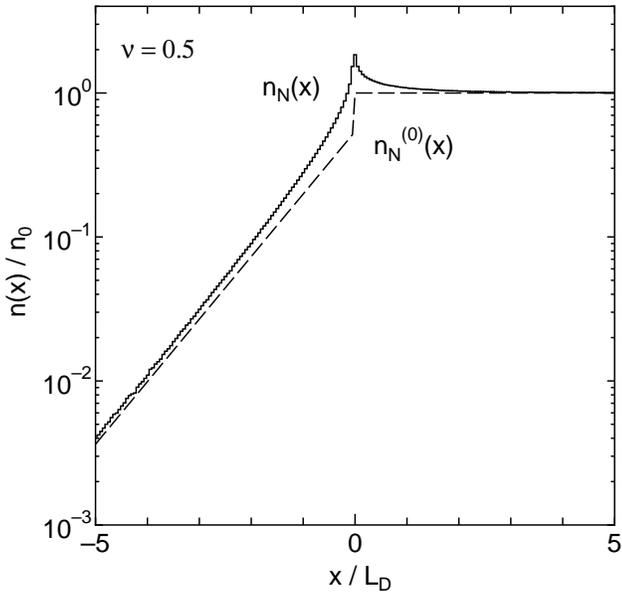}
  \caption{
    The density of the scattering points for no-boundary solution 
    $\nN(x)$ (solid histogram) for $\nu=0.5$ obtained by a 
    Monte Carlo simulation 
    and the asymptotic approximation $\nNz(x)$ (dashed curve). 
  }
  \label{fig:nNx}
\end{figure}

\begin{figure}
  \leavevmode
    \epsfysize=8cm
    \epsfbox{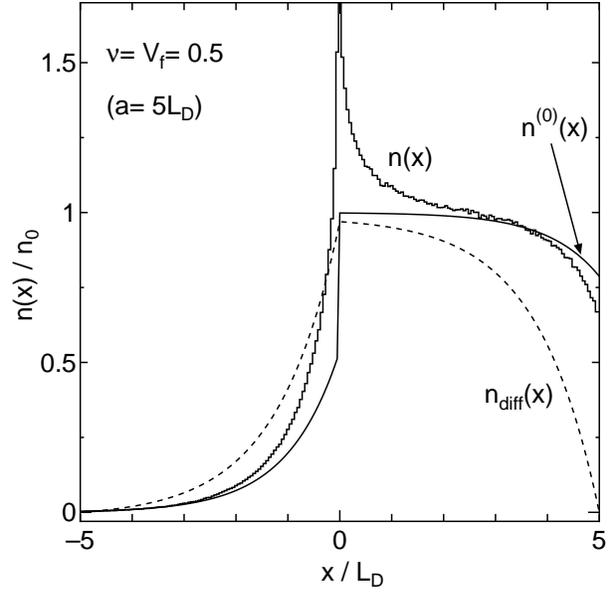}
  \caption{
    The density of the scattering points $n(x)$ (solid histogram)
    obtained by a Monte Carlo simulation is shown for 
   $\nu=\Vf=0.5, a=5\LD, b\to\infty$
    together with our approximate solution $n^{(0)}(x)$ and 
    that of the conventional diffusion equation $n_{\rm diff}(x)$ 
   (dotted curve).
  }
  \label{fig:nx}
\end{figure}

\Reffig{nNx} shows an example of the no-boundary solution 
$\nN(x)$ (solid histogram) for $\nu = 0.5$,
which is obtained by the Monte Carlo simulation,
together with the asymptotic approximation $\nNz(x)$ 
(\Refeq{approx_nN}; dashed curve).
It confirms that
$\nNz(x)$ approaches $\nN(x)$ asymptotically for 
$x \rightarrow -\infty$ or $x \rightarrow +\infty$.
Although $\nN(x)$ has a peak around $x=0$,
our approximation ignores this peak as 
described in \Refsec{approximate solutions}.
\Reffig{nx}
shows an example of the results when
the single boundary
is located at $a = 5\LD (b \to \infty)$. The density of scattering points 
$n(x)$ (solid histogram) for $\nu = 0.5$ is obtained 
by a Monte Carlo simulation, along with our approximate 
solutions and that of the conventional diffusion equation
(Drury 1983; his equation (3.41))
$n_{\rm diff}(x)$ (dotted curve).
It is clearly seen that the conventional diffusion equation does not 
give a sufficiently good approximation to the real solution 
for high $\nu$.
Especially, the random walk result shows that 
$n(x)$ has a finite value at the absorption barriers,
while $n_{\rm diff}(x)$ becomes $0$ there.
The small difference between the Monte Carlo result
and the our approximation
near the absorbing boundary
is mainly caused by
omitting the peak in $\nN(x)$ around $x=0$
in making our approximation.

\subsection{Return probability densities}
Here, we present the return probabilities of particles 
for the isotropic large angle scattering model
under the approximation made in the previous section. 
This model gives
\begin{equation}
\gp(l) = \frac{l}{2} \left\{ \frac{\Lp}{l} - \ln(1+\frac{\Lp}{l}) \right\},
\end{equation}
\begin{equation}
\gm(l) = \frac{l}{2} \left\{ \frac{\Lm}{l} - \ln(1+\frac{\Lm}{l}) \right\},
\end{equation}
\begin{equation}
\hm = \frac{\Lm^2}{4}
\end{equation}
and
\begin{equation}
\label{eq:const_A_iso}
A = \left(\frac{\Lm}{\Lp}\right)^2 \frac{1}{\psi(\Rp)}
\end{equation}
where we define
\begin{equation}
\label{eq:psi}
\psi(x) := 2 \frac{x - \ln(1+x)}{x^2}.
\end{equation}
From these quantities,
the probability densities of pitch angle at return $\Pm$ and $\Pp$ 
are calculated by \Refeq{approx_Pm} and \refeq{approx_Pp},  
while those for the multi-step approximation are calculated by 
\Refeq{approx_Pm_m} and \refeq{approx_Pp_m}, 
respectively.
The total return probability $\Pret$ is also calculated 
by \Refeq{approx_Pret} or \Refeq{approx_Pret_m}.

\begin{figure*}
  \leavevmode
    \epsfysize=8cm
    \epsfbox{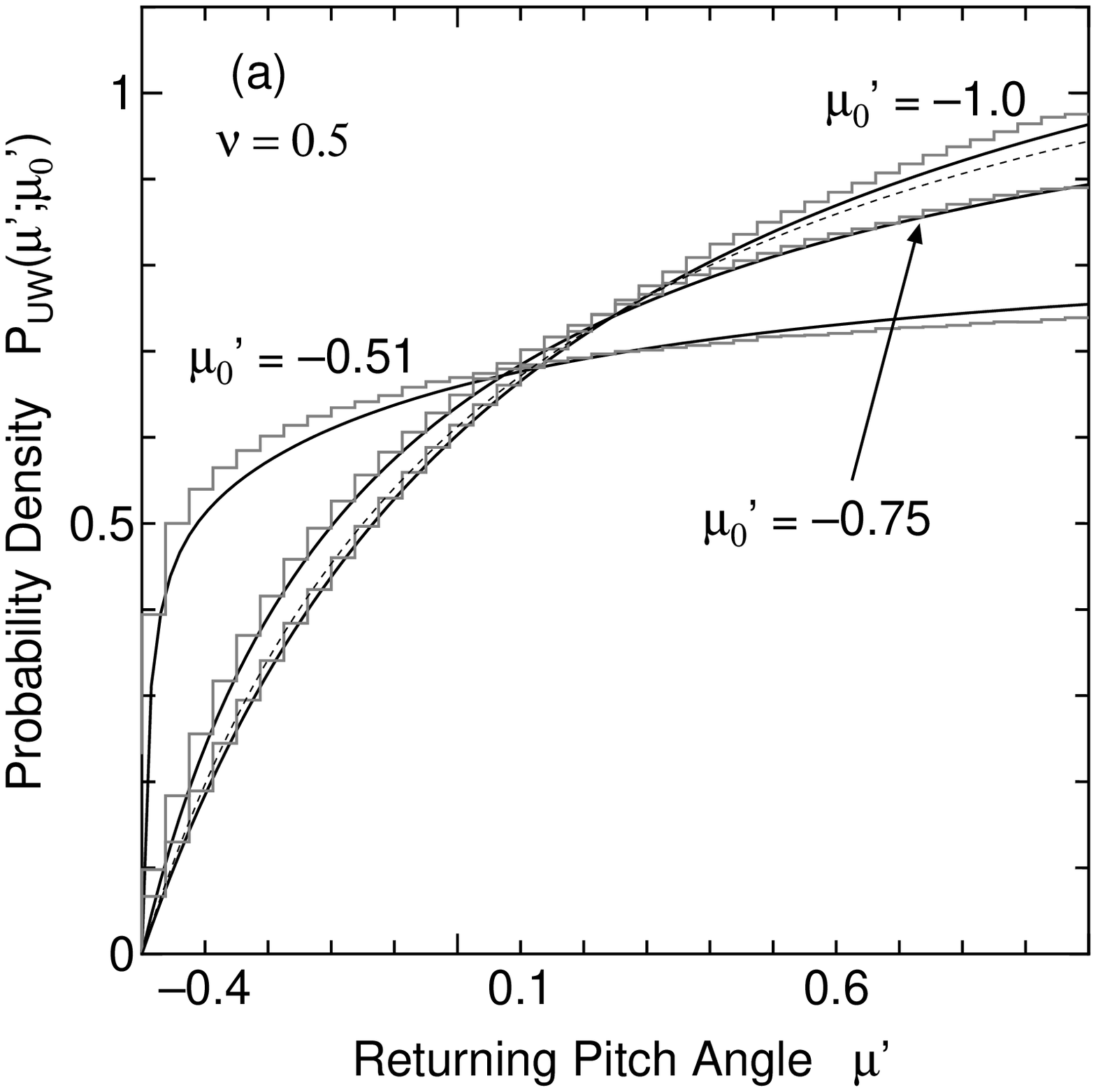}
    \epsfysize=8cm
    \epsfbox{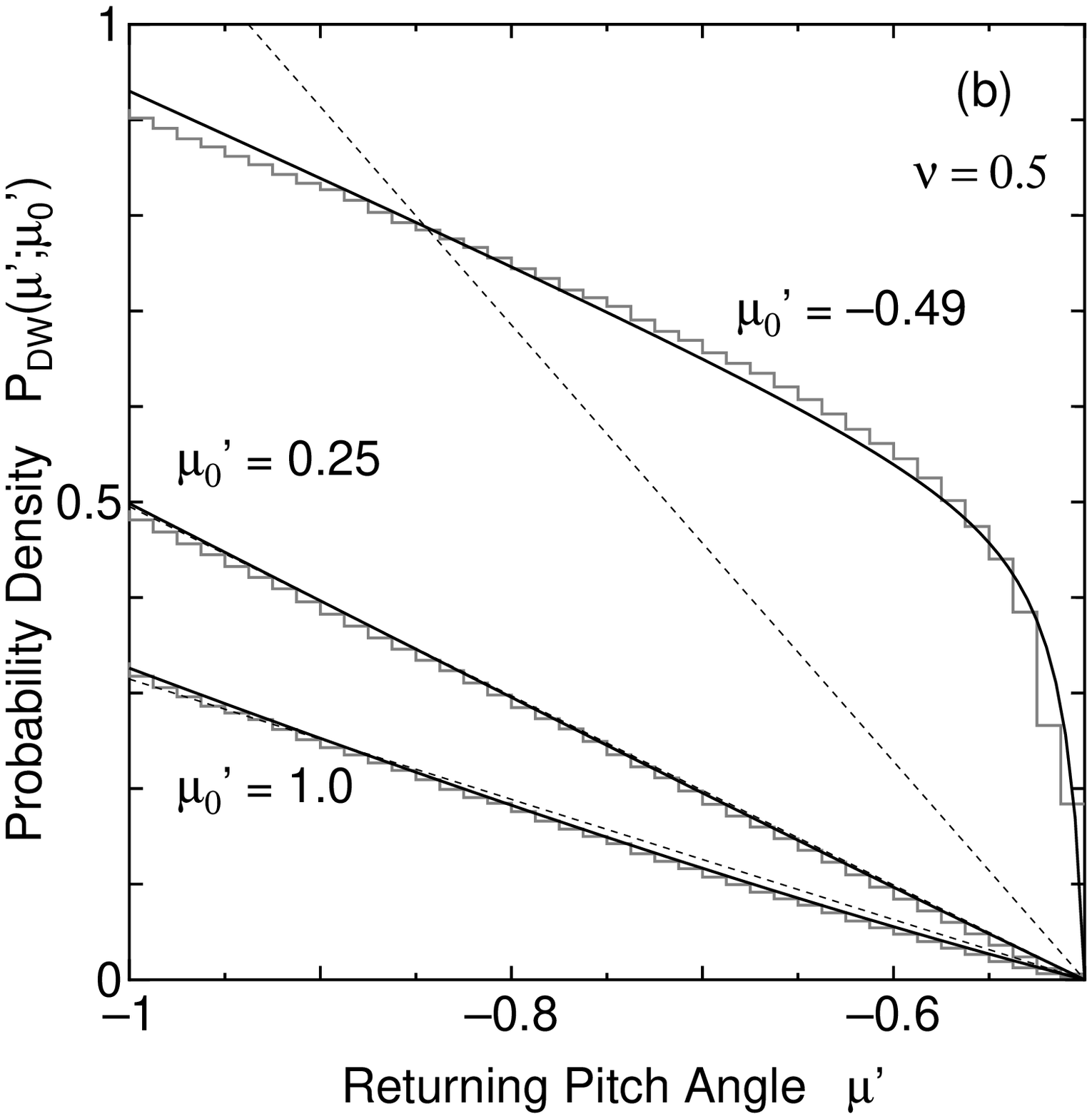}
  \caption{
    The probability densities of pitch angle at return 
    for $\nu = 0.5$ (a) $\Pm(\mu';\mu_0')$ and (b) $\Pp(\mu';\mu_0')$
    for several values of the initial pitch angle $\mu_0'$. 
    Results of our approximation (solid curves),
    the multi-step approximation (dotted curves)
    and Monte Carlo simulation (grey histograms) are shown.
  }
  \label{fig:p}
\end{figure*}
\Reffig{p}
shows the probability density of pitch angle at return 
$\Pm(\mu_0')$ in (a) and $\Pp(\mu_0')$ in (b) 
for $\nu=0.5$ and for several values of the initial pitch angle 
cosine $\mu_0'$. 
Our approximation (solid curves) gives a fairly good fit
to the Monte Carlo simulation (grey histograms)
for all initial pitch angle cosine $\mu_0'$.
For small $\mu_0'$, the fit deviates a little
because the effects of the peak in $\nN(x)$ around $x=0$,
which is ignored in our approximation,
affect the probability density.
The multi-step approximation (dotted curves)
does not fit well when the initial mean free path $\lmdz$ is small
because effects of the return after only a few steps of scattering 
becomes important. But, for large $\lmdz$ and for $\Pp$,
the multi-step approximation can fit to the Monte Carlo simulation.

\begin{figure}
  \leavevmode
    \epsfysize=8cm
    \epsfbox{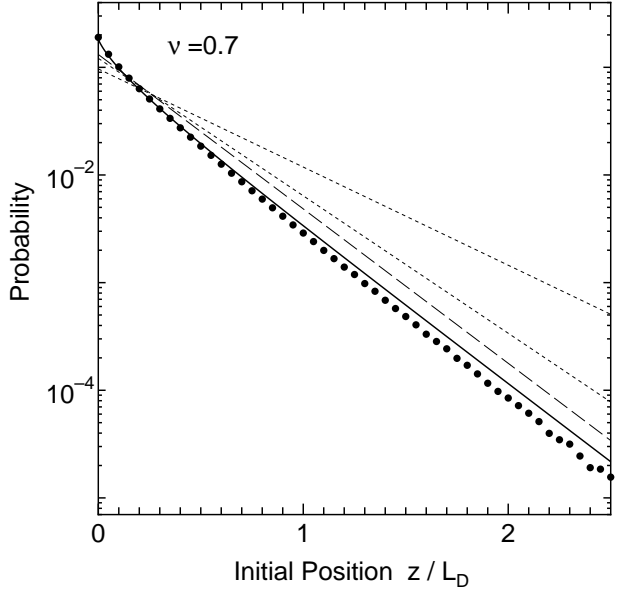}
  \caption{
    The absorption probability from $z$, $\Pabs(z)$, for $\nu=0.7$.
	Results of our approximation (solid curve),
    the multi-step approximation (dashed curve),
    and Monte Carlo simulation (filled circle) are represented.
    The results using
    the length $1/(3\Gf\nu)$ (lower dotted curve)
    or $1/(3\nu)$ (upper dotted curve)
    instead of $\LD$ in the multi-step approximation
    are also shown as in Fig.\~2.
    The curve $1/(3\nu)$ also corresponds to
    the extrapolation of Peacock's expression (his eq.(24))
    to relativistic fluid speed.
  }
  \label{fig:Pabs}
\end{figure}

The expression of the absorption probability $\Pabs(b)$ 
by our approximation
is a little cumbersome and is presented in \Refapn{Pabs}.
The absorption probability by the multi-step approximation
is given by \Refeq{Pabs_m} together with \Refeq{const_A_iso}.
\Reffig{Pabs}
presents $\Pabs(b)$
as a function of the initial position $b$ for $\nu = 0.7$.
The result of our approximation (solid curve) gives
a good fit to the Monte Carlo results (filled circles)
(small difference is in part caused by omitting the peak in $\nN(x)$).
The multi-step approximation (dashed curve)
deviates from the Monte Carlo results,
but gives a correct scale length ($\LD$). 
This result directly confirms that
the diffusion length for such large $\nu$
is given by our diffusion length $\LD$.
Results using the conventional diffusion length $\LDz$
instead of $\LD$
in the expression of the multi-step approximation
(\Refeq{Pabs_m} and \refeq{const_A})
for two case as in \Reffig{scale}
are also represented;
lower dotted curve for $1/(3\Gf\nu)$
and upper dotted curve for $1/(3\nu)$).
It should be noted that
if we extrapolate the expression (24) in Peacock (1981),
which was derived by adopting the diffusion approximation
for particles in downstream of the shock front,
to relativistic fluid speed in fluid frame
and then transform it to the expression for the boundary rest frame,
it turns out to correspond to the multi-step approximation \refeq{Pabs_m}
except that it has the scale length of return $1/(3\nu)$ (in our unit).
These results do not give a satisfactory fit.

\begin{figure}
  \leavevmode
    \epsfysize=8cm
    \epsfbox{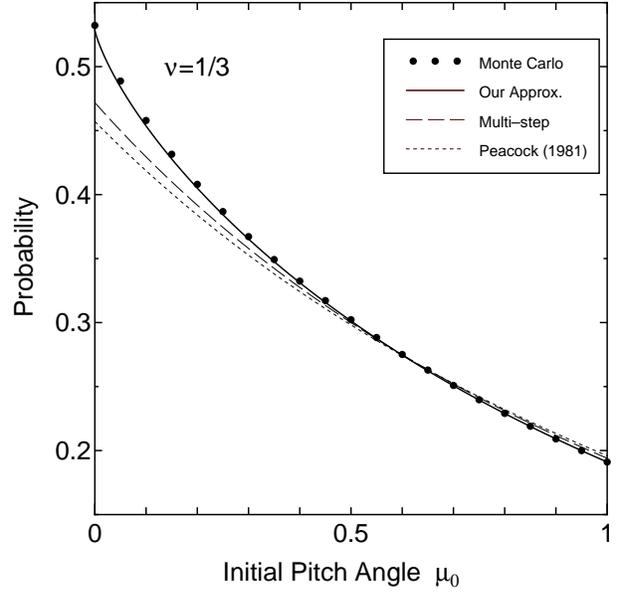}
  \caption{
    The total return probability $\Pret(\mu_0)$ for $\nu = 1/3$ 
    from the downstream.
    Here, the initial pitch angle cosine $\mu_0$ is 
    measured in the boundary rest frame.
    Results of our approximation (solid curve),
    the multi-step approximation (dashed curve),
    Peacock's approximation (dotted curve)
    and Monte Carlo simulation (filled circle) are shown.
  }
  \label{fig:pret}
\end{figure}

\Reffig{pret}
presents
the total return probability $\Pret(\mu_0)$
from the downstream 
for mildly high $\nu$ ($\nu = 1/3$)
as a function of initial pitch angle cosine $\mu_0$,
which is measured in the boundary rest frame.
The results of our approximation (solid curve)
excellently agree with the Monte Carlo simulation (filled circles).
The dotted curve shows the approximate solution of \Peacock81 (shown in his fig.2).
The result of the multi-step approximation (dashed curve),
which uses the correct length $\LD$,
is similar to Peacock's one,
but slightly better than it.
These two diffusive approximations do not agree with
the Monte Carlo results for small $\mu_0$.
The reason is similar to that for $\Pm$ and $\Pp$;
in diffusive approximations, effects of return after only a few steps 
of scattering are not accounted for.
In general, for higher $\nu$ or smaller $\lmdz$,
such effects become important
and these diffusive approximations become worse.

\subsection{Density of scattering points}
\begin{figure*}
  \leavevmode
    \epsfysize=8cm
    \epsfbox{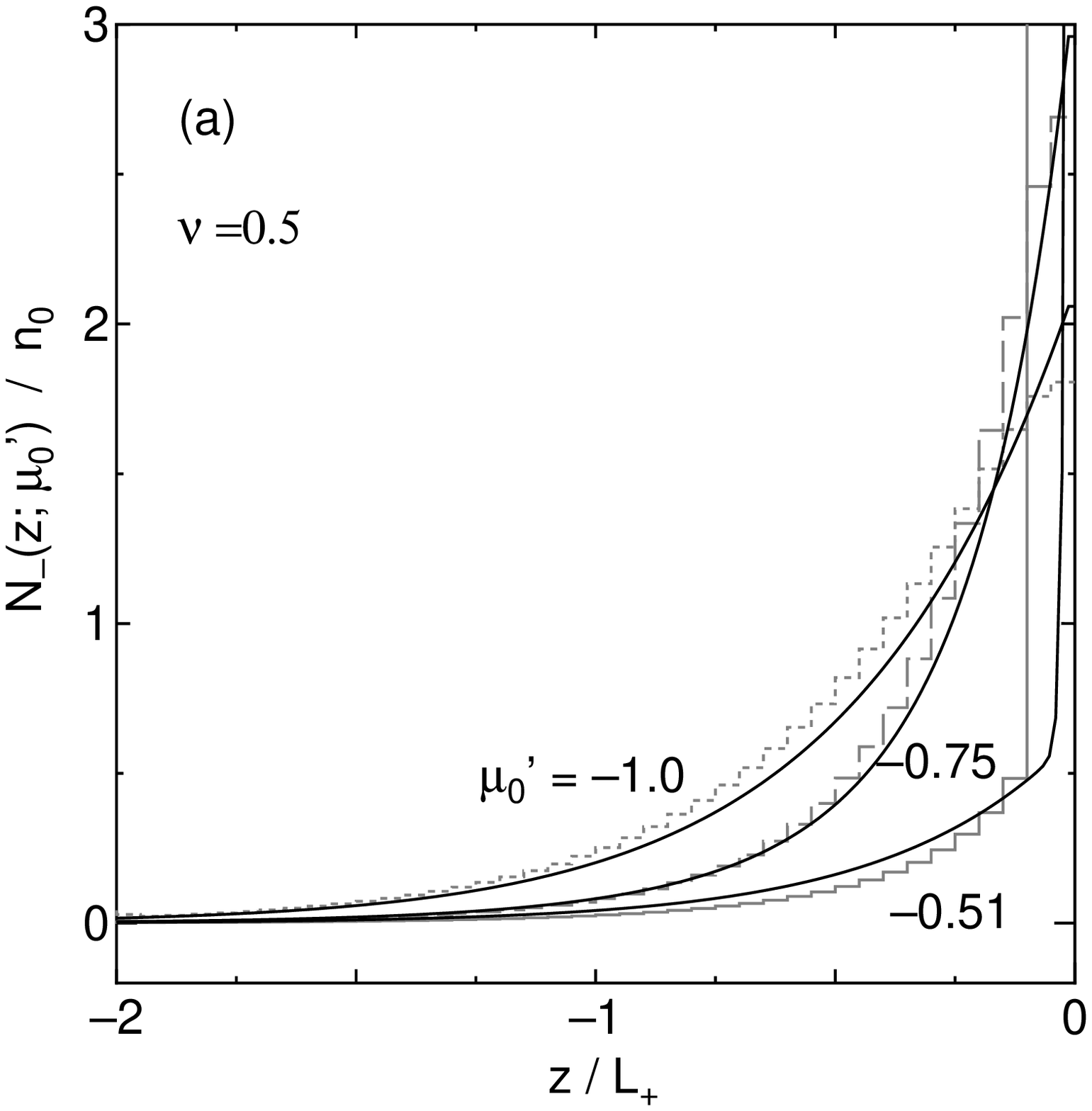}
    \epsfysize=8cm
    \epsfbox{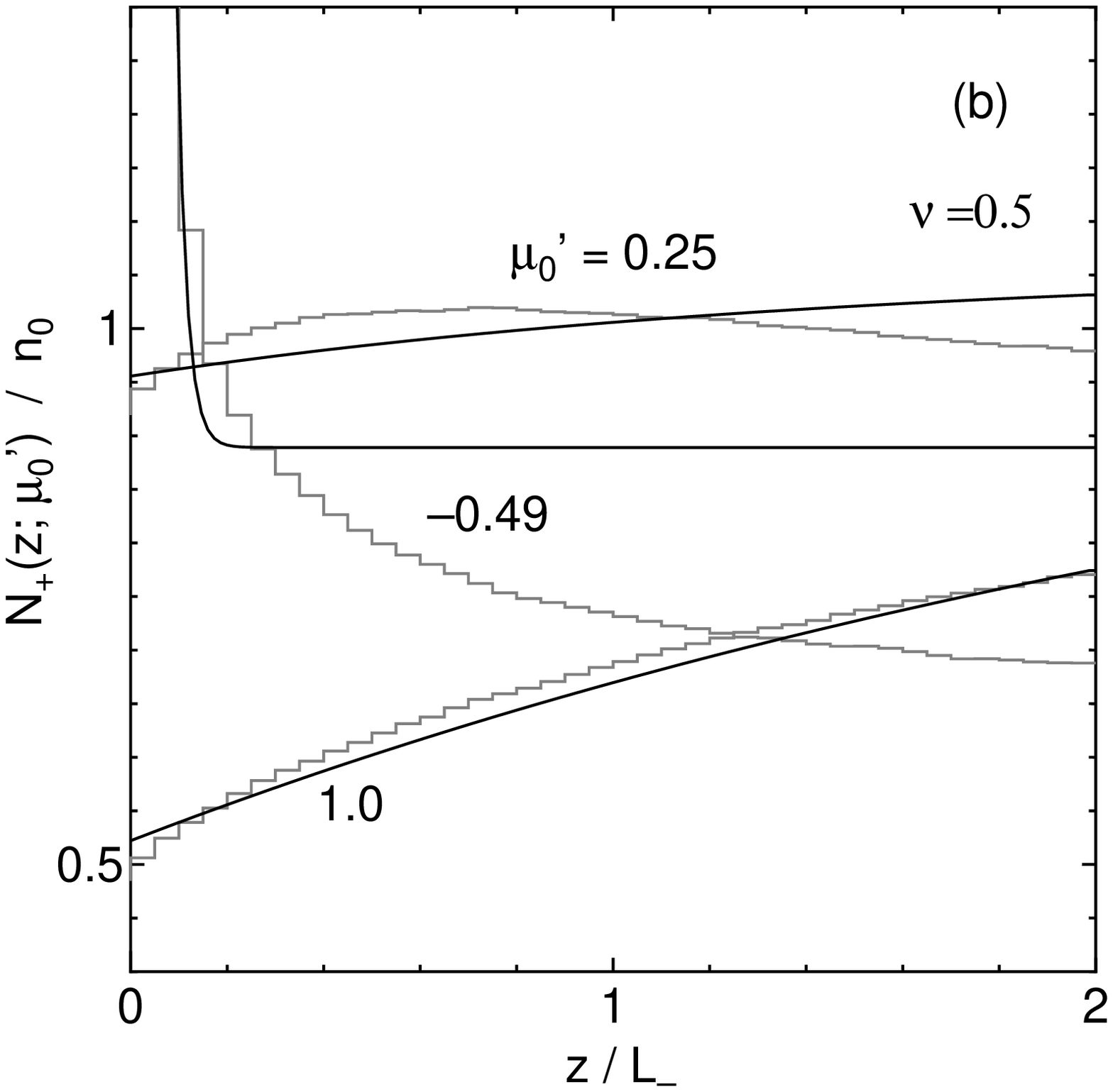}
  \caption{
    The density of the scattering points for particles which 
    cross the boundary with the initial pitch angle $\mu_0'$,    
    (a) $\Nm(z;\mu_0')$ and (b) $\Np(z;\mu_0')$. 
    Results of our approximation (solid curves),
    and Monte Carlo simulation (histograms) are shown
    for $\nu = 0.5$.
  }
  \label{fig:Nx}
\end{figure*}

Here, we show the results of the density of the scattering 
points for the particles which cross the shock front 
with the pitch angle cosine $\mu_0'$.
\Reffig{Nx}
shows $\Nm(z;\mu_0')$ in (a) and $\Np(z;\mu_0')$
in (b) calculated by the Monte Carlo simulation (histograms) 
for several values of the initial pitch angle $\mu_0'$ 
along with our approximate solutions.
It is seen that our approximation (solid curves)
coincides well with Monte Carlo simulation. 
But, a discrepancy is recognized for $z > 0.3 \Lm$ 
in (b). 
Because the normalization conditions used in our approximation
(the \Refeq{total_Pm}, \refeq{Np_condition})
employ only the information of finally absorbed particles
and do not care about escaping particles to the far downstream, 
our approximation fails to give a good fit in a distant region 
where escaping particles dominate, while it 
gives a satisfactory fit in a near region
where almost all the last scattering points exist.
This also means that our approximation works well
for absorbed particles,
therefore
the return probability densities themselves are 
represented by our approximations quite well
as shown in \Reffig{p} and \Reffig{pret}.

\begin{figure*}
  \leavevmode
    \epsfysize=8cm
    \epsfbox{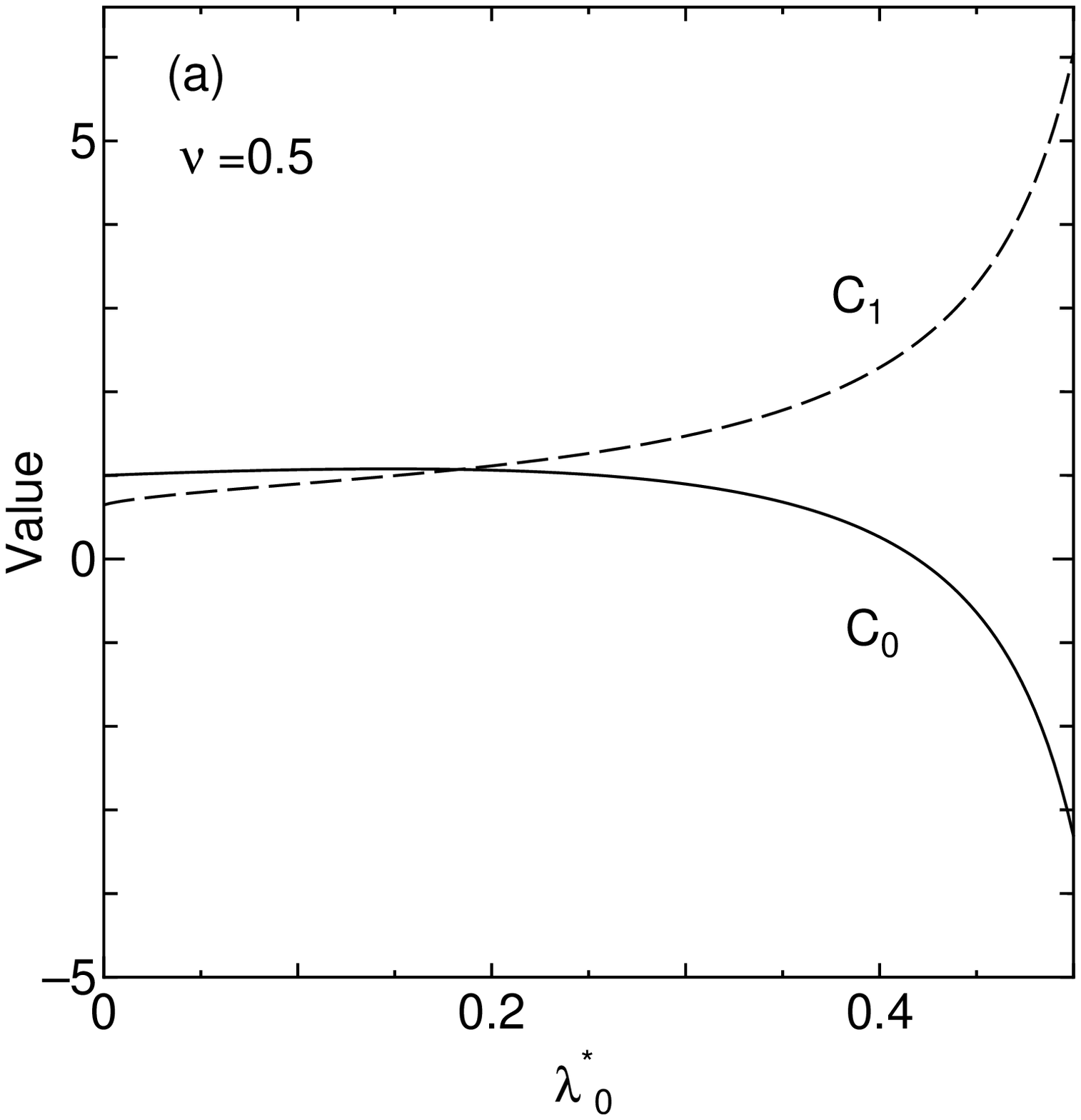}
    \epsfysize=8cm
    \epsfbox{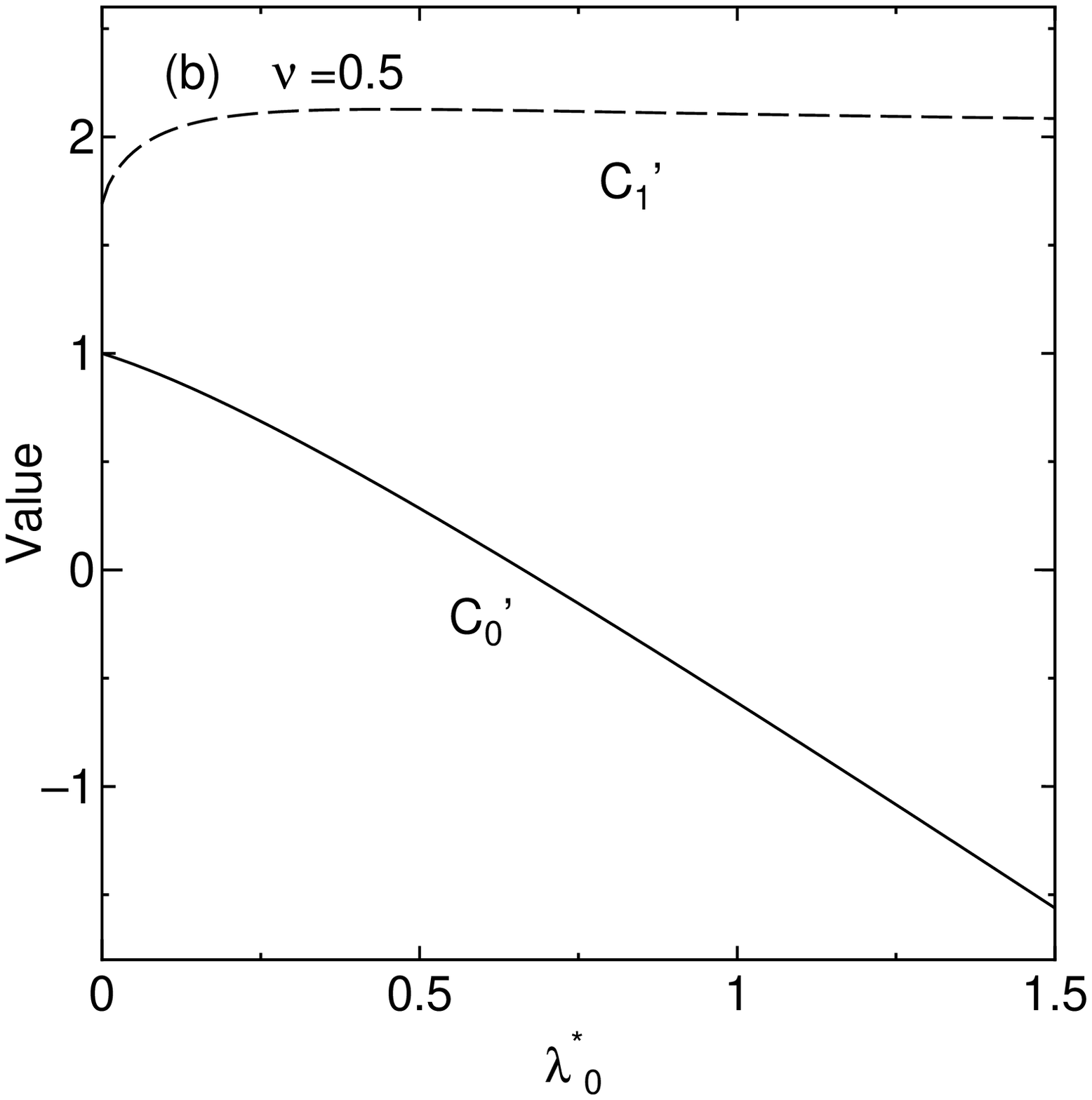}
  \caption{
    The coefficients in the expression of
    (a) $\Nm(z;\mu_0)$ and (b) $\Np(z;\mu_0)$
    for $\nu = 0.5$
  	are shown as functions of the initial mean free path $\lmdz$.
  }
  \label{fig:C0_C1}
\end{figure*}

Another important feature of $\Np$ and $\Nm$ is that 
the density of the scattering points near the boundary
depends on the initial pitch angle $\mu_0'$.
Here, we explain this feature for $\Np$ (that for 
$\Nm$ is similar).
\Reffig{Nx} (b) shows that 
$\Np(z)$ increases when $z \rightarrow 0$ for small $\lmdz$ 
but that it decreases when $z \rightarrow 0$ for large $\lmdz$.
This can be explained as already mentioned in Section 3;
for small $\lmdz$
the scattering point of few step particles are important,
while for large $\lmdz$
the effect of finite initial mean free path is important.
In our approximate expression of \refeq{approx_Np},
while $C_1'$ is always positive,
$C_0'$ becomes negative for large $\lmdz$,
as shown in \Reffig{C0_C1} (b).
(This feature also appears for $\Nm$, as shown in \Reffig{C0_C1} (a).)
Because of this feature, when we average the return probability over 
an incoming particle distribution,
the first term in \refeq{approx_Np},
which denotes the non-diffusive effects,
will become small enough
compared with the diffusive term which is associated with $C_1'$.
Then, there appears a possibility to use 
the multi-step approximation to determine the {\it averaged}
pitch angle distribution over all incident particles 
even if $\nu$ is large.
The same consideration applies for $\Nm$, too. 
As will be shown in the following section,
at least for the isotropic large angle scattering model and
for typical incoming $\mu_0'$ distributions,
{\it this is actually the case}. 
In practice, for this reason,
we can use the multi-step approximation
to determine the pitch angle distribution at the shock front
for {\it arbitrary} shock speed
as shown in the following section.

\section[]{Application to the Shock\\* Acceleration}
\label{sec:shock acceleration}
In this section, we apply the random walk theory developed so far 
to the shock acceleration and calculate the spectral index 
of the accelerated particles for 
the isotropic large angle scattering model described in the previous section.
In the following,
we consider the acceleration in parallel shocks
where the fluid speed measured in the shock rest frame
are uniform on each side of the shock front;
$\Vu$ for the upstream and $\Vd$ for downstream respectively.
We also consider only relativistic particles 
($v' \sim 1$, i.e., $\nu \sim \Vf$)
because, in this case,
the ratio $\nu$ does not change
on each side of the shock front
even if the particle gains energy
in the process of multiple crossing at the shock front.
In addition, for such particles,
the Lorentz transformation of $\mu$ to $\mu'$ becomes as simple as
\begin{equation}
\label{eq:mu_trans}
\mu' = \frac{\mu - \Vf}{1 - \mu \Vf},
\quad
d\mu' = \frac{1 - \Vf^2}{(1-\mu\Vf)^2} d\mu, 
\end{equation}
where $\mu$ is 
the pitch angle cosine measured in the shock rest frame.
Hereafter,
we use subscripts u, d or superscripts (u), (d)
to indicate quantities in the upstream and downstream regions, 
respectively. 
The pitch angle cosine measured in the upstream fluid frame is denoted 
by $\muu$ and that 
measured in the downstream fluid frame by $\mud$.

In the following,
we present the results only for two simple jump conditions 
$r=\Vu/\Vd=4$ and $r=3\Vs^2$ (i.e., $\Vu\Vd = 1/3$).
But, since 
the results are expressed in terms of $\Vu$ and $\Vd$, 
they are applicable to any jump conditions. 
(Jump conditions for relativistic shocks are
discussed in Blandford \& McKee \shortcite{BM76},
\Peacock81, Appl \& Camenzind \shortcite{AC88}
and Heavens \& Drury \shortcite{HD88}.)

\subsection[]{Method of determining the pitch angle\\* distribution at the shock front}
The spectral index of the accelerated particles $\sigma$
is given by
\begin{equation}
\label{eq:index}
\sigma = 1 - \frac{\ln(\PRET)}{{\lnd}}
\end{equation}
where $\PRET$ is the averaged return probability for the particles
crossing the shock front toward downstream and 
$\lnd$ is the logarithm of the energy gain factor
 (see Peacock 1981).
(Here, $\langle \rangle$ denotes to average over the 
pitch angle distribution crossing the shock front.)
$\lnd$ is given as
\begin{equation}
\label{eq:lnd}
\lnd = 2 \ln \Grel + \lndud + \lnddu,
\end{equation}
where
\begin{equation}
\lndud := \left< \ln(1 + \Vrel \muuud) \right>,
\end{equation}
\begin{equation}
\lnddu := \left< \ln(1 - \Vrel \muddu) \right>
\end{equation}
and $\Vrel$ is
the relative velocity of the upstream fluid
with respect to the downstream fluid,
\begin{equation}
\Vrel = \frac{\Vu - \Vd}{1 - \Vu\Vd},
\qquad \Grel := \frac{1}{\sqrt{1-\Vrel^2}}.
\end{equation}
If the dependences on $\mu'$ and $v'$ in $\lambda$ and 
$\Pmu$ are separable,
these quantities are independent of the particle energy, i.e., $v'$.

Introducing the normalized pitch angle distribution of particles 
crossing the shock front
from upstream toward downstream by $\pud(\muu)$, and
that from downstream toward upstream by $\pdu(\mud)$,
i.e., 
\[
\int_{-\du}^1 \pud(\muu) d\muu = 
\int_{-1}^{-\dd} \pdu(\mud) d\mud = 1,
\]
we can write relevant quantities as follows.
Using $\Pretd(\mud)$ and $\pud(\mud)$,
$\PRET$ is written as
\begin{equation}
\label{eq:PRET}
\PRET = \int_{-\dd}^1 \Pretd(\mud) \pud(\mud) d\mud.
\end{equation}
The logarithm of the energy gain factor
is given by
\begin{equation}
\label{eq:lndud}
\lndud 
= \int_{-\du}^1 \! \ln(1+\Vrel \muuud)
    \; \frac{\Pretd(\mudud)}{\PRET} \pud(\muuud) d\muuud
\end{equation}
and
\begin{equation}
\lnddu 
= \int_{-1}^{-\dd} \! \ln(1 - \Vrel \muddu) \; 
\pdu(\muddu) d\muddu.
\end{equation}
The factor $\Pretd(\mudud)/\PRET$ in \Refeq{lndud} means that 
only particles which return again to the shock front are to be 
considered.

The pitch angle distribution $\pud(\muu)$ and $\pdu(\mud)$
in a steady state are determined as follows.
If $\pud(\muu)$ is known, 
$\pdu(\mud)$ is determined by $\pud(\muu)$,
\begin{equation}
\label{eq:pdu}
\pdu(\mud)
= \int_{-\du}^{1} \frac{\Pdwd(\mud ; \mudz)}{\Pretd(\mudz)} 
\pud(\muuz) d\muuz.
\end{equation}
On the other hand,
$\pud(\muu)$ is also determined by $\pdu(\mud)$,
\begin{equation}
\label{eq:pud}
\pud(\muu) = \int_{-1}^{-\dd} \Puwu(\mu ; \mu_0) \pdu(\mudz) d\mudz.
\end{equation}
Therefore, in a steady state,
$\pdu(\mud)$ satisfies an integral equation
(a Fredholm equation of second kind)
\begin{equation}
\label{eq:integral_eq2}
\pdu(\mud) = \int_{-1}^{-\dd} K(\mud; \mudz) \pdu(\mudz) d\mudz,
\end{equation}
where the kernel of this integral equation is given by 
\begin{equation}
\label{eq:kernel2}
K(\mud ; \mudz)
:=
\int_{-\du}^{1} \frac{\Pdwd(\mud; \mudo)}{\Pretd(\mudo)}
            \Puwu(\muuo; \muuz) d\muuo.
\end{equation}
This kernel denotes the probability density of the particle
which crosses the shock front from downstream to upstream 
with $\mudz$
crosses the shock front again from downstream to upstream with $\mud$
after one cycle.
If $\pdu(\mud)$ can be solved,
$\pud(\muu)$ is calculated through \Refeq{pud}.
Although in the above
we have made the integral equation for $\pdu(\mud)$,
one can alternatively make the integral equation for $\pud(\muu)$. 

\subsection[]{Approximate solution of the pitch angle\\* distribution at the shock front}
It is generally difficult to solve
the integral equation \refeq{integral_eq2} analytically
even if our approximation are used.
Then, we numerically evaluate the kernel 
for our approximation. 
On the other hand,
in the multi-step approximation,
$\pud(\mu)$ and $\pdu(\mu)$ can be calculated analytically
without solving the integral equation \refeq{integral_eq2}.
Because the dependence of $\Pdw(\mu';\mu'_0)$ in 
\Refeq{approx_Pp_m} are separable in $\mu'$  
and $\mu'_0$ in this approximation,
it becomes 
\begin{equation}
\label{eq:pdu_D}
\pdu(\mud)
= \frac{\Pmu \lmdd}{\hm}.
\end{equation}
Because $\Puw(\mu';\mu'_0)$ in \Refeq{approx_Pm_m}
is independent of $\mu'_0$, we obtain 
\begin{equation}
\label{eq:pud_D}
\pud(\muu) = \frac{1}{\gp(\LD)} \frac{\Pmu \lmd}{\lmd + \LD}.
\end{equation}
These results agree with the results obtained by Peacock(1981),
but $\LD$ was replaced by $\LDz$ in \Refeq{pdu_D}.

\begin{figure}
  \leavevmode
    \epsfysize=8cm
    \epsfbox{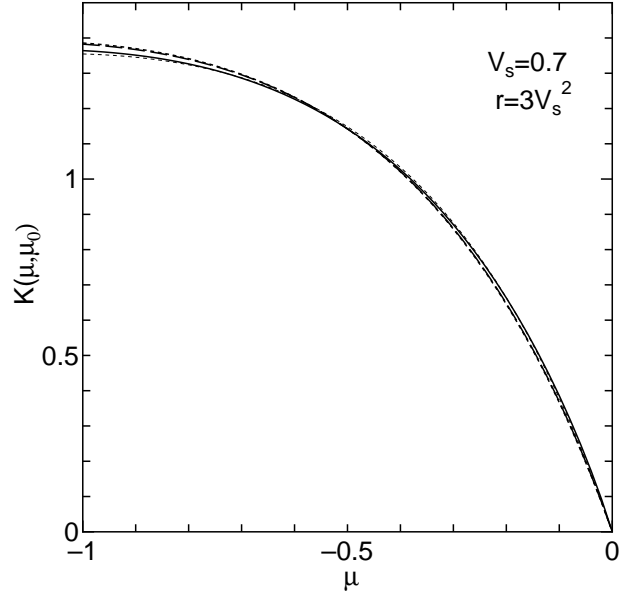}
  \caption{
    The kernel of integral equation (100) 
    for $\Vs = 0.7$ and $r=3 \Vs^2$. 
    The numerical evaluation based on our approximation
    for $\mu_0 = -0.01$ (solid curve), $-0.5$ (dashed curve) and $-1.0$ 
    (dotted curve)
    are shown together with the result of the multi-step approximation 
    (thin dotted curve).
    The results turn out to be almost independent of $\mu_0$ and 
    all curves give similar results.}
  \label{fig:kernel}
\end{figure}
\Reffig{kernel}
shows the kernel of \Refeq{integral_eq2}
for $\Vs = 0.7$ and $r=3\Vs^2$.
The numerical evaluation based on our approximation for 
$\mu_0 = -0.01$ (solid curve),
$-0.5$ (dashed curve) and $-1.0$ (dotted curve)
together with that of the multi-step approximation are shown.
It is seen that the kernel is almost independent of the initial 
pitch angle cosine $\mu_0$.
The reason for this coincidence is considered to be due to 
compensation effects of several non-diffusive effects
when averaged over the initial pitch angle
as was mentioned in the previous section.
Although this is partly due to our assumption of 
the isotropic large angle scattering, 
it is of a great importance in applications to the 
shock acceleration. 
Although the multi-step approximation is a poor approximation 
to the detailed particle transport, it can be a good approximation 
when we average over the pitch angle distribution. 
Thus, this result gives a justification for using the multi-step 
approximation even for relativistic shocks. 

\begin{figure*}
  \leavevmode
    \epsfysize=11cm
    \epsfbox{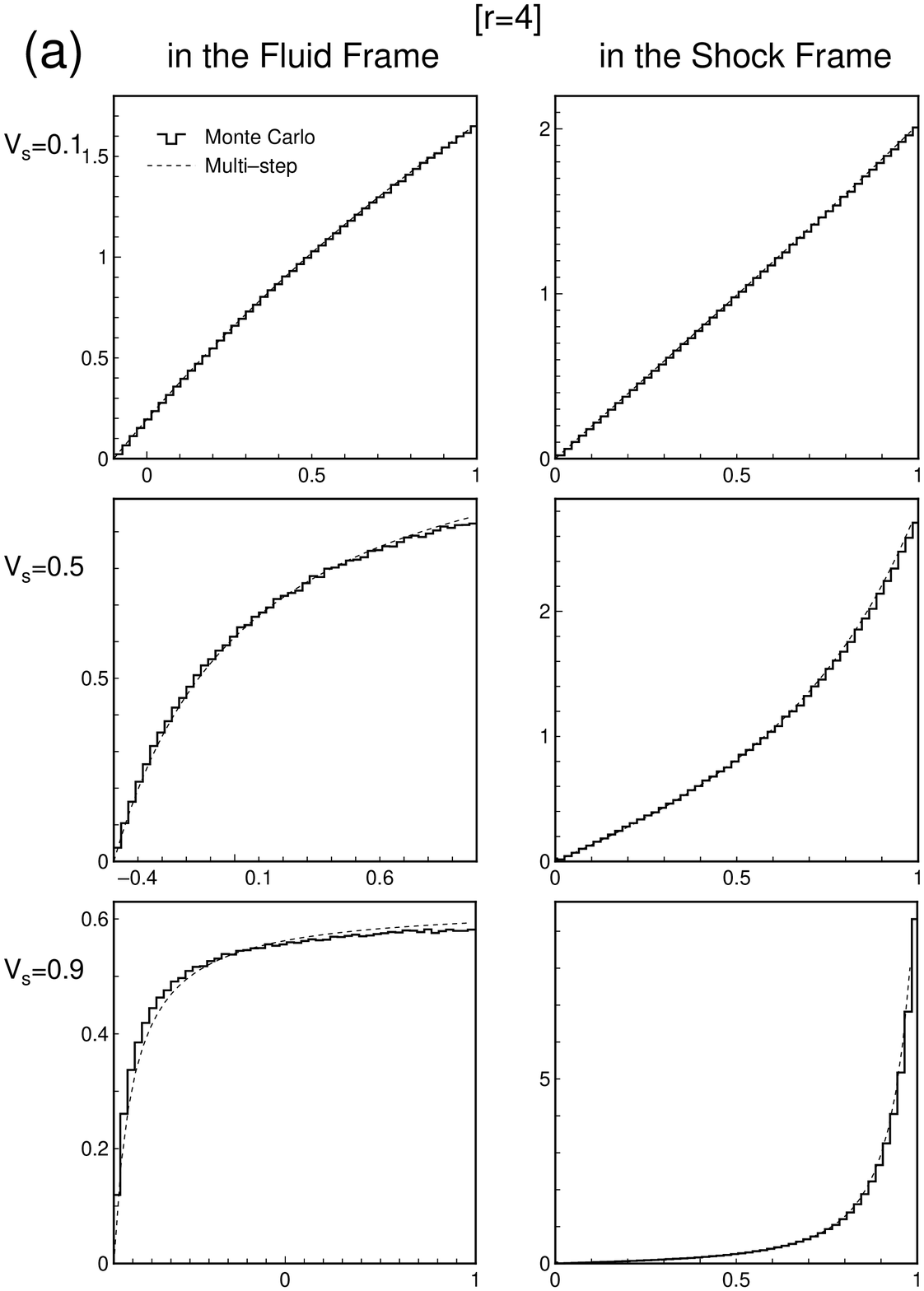}
    \epsfysize=11cm
    \epsfbox{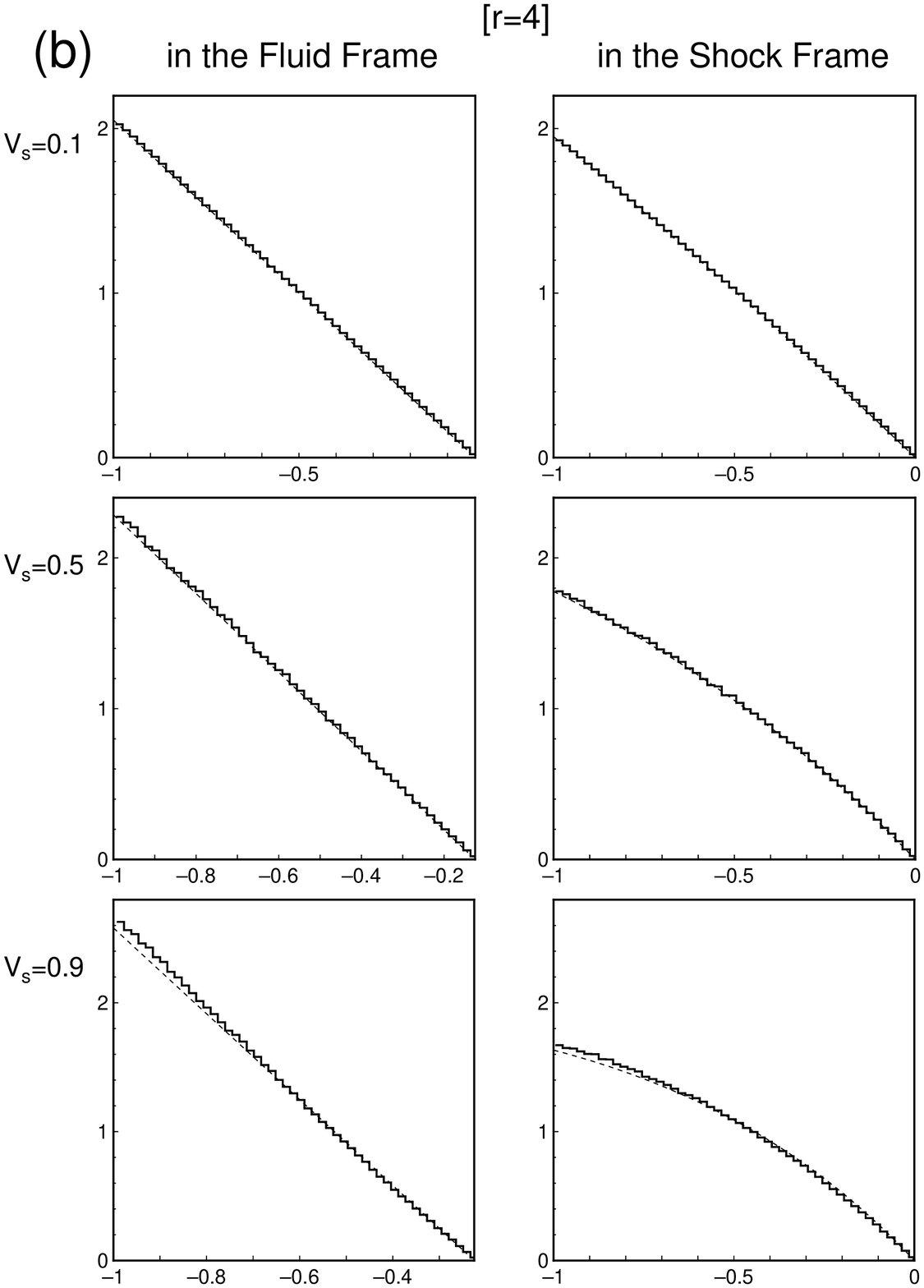}
  \caption{
    The pitch angle distribution of the particles crossing the shock front. 
    Figure (a) refers to the particles crossing from 
    upstream to downstream while figure (b) does those 
    from downstream to upstream.
    The pitch angle distribution both
    in the fluid frame and the shock rest frame are shown.
    The results of the multi-step approximation (dotted curve)
	and the Monte Carlo simulation (solid histogram) are shown.
  }
  \label{fig:pa_dist}
\end{figure*}

\Reffig{pa_dist}
presents the pitch angle distribution at the shock front 
in various frames for various shock velocities.
The results of the multi-step approximation (dotted curve)
and the Monte Carlo simulation (solid histogram) are shown.
As is seen, the results of the multi-step approximation
agree  quite well with the Monte Carlo results
even for highly relativistic shocks.
This is not surprising once we admit that the kernel is well 
approximated by the multi-step approximation. 
Thus, we have confirmed the point mentioned above that 
the multi-step approximation gives a good fit to 
the steady state pitch angle distribution at the shock for 
any shock speed. 
Also,
it is to be noted that
the assumptions used in Peacock (1981)
are adequate for arbitrary shock speed
if the correct diffusion length $\LD$ is used
in the downstream as in the multi-step approximation.

\subsection{Approximate solution of the spectral index}
\begin{figure*}
  \leavevmode
    \epsfysize=7.95cm
    \epsfbox{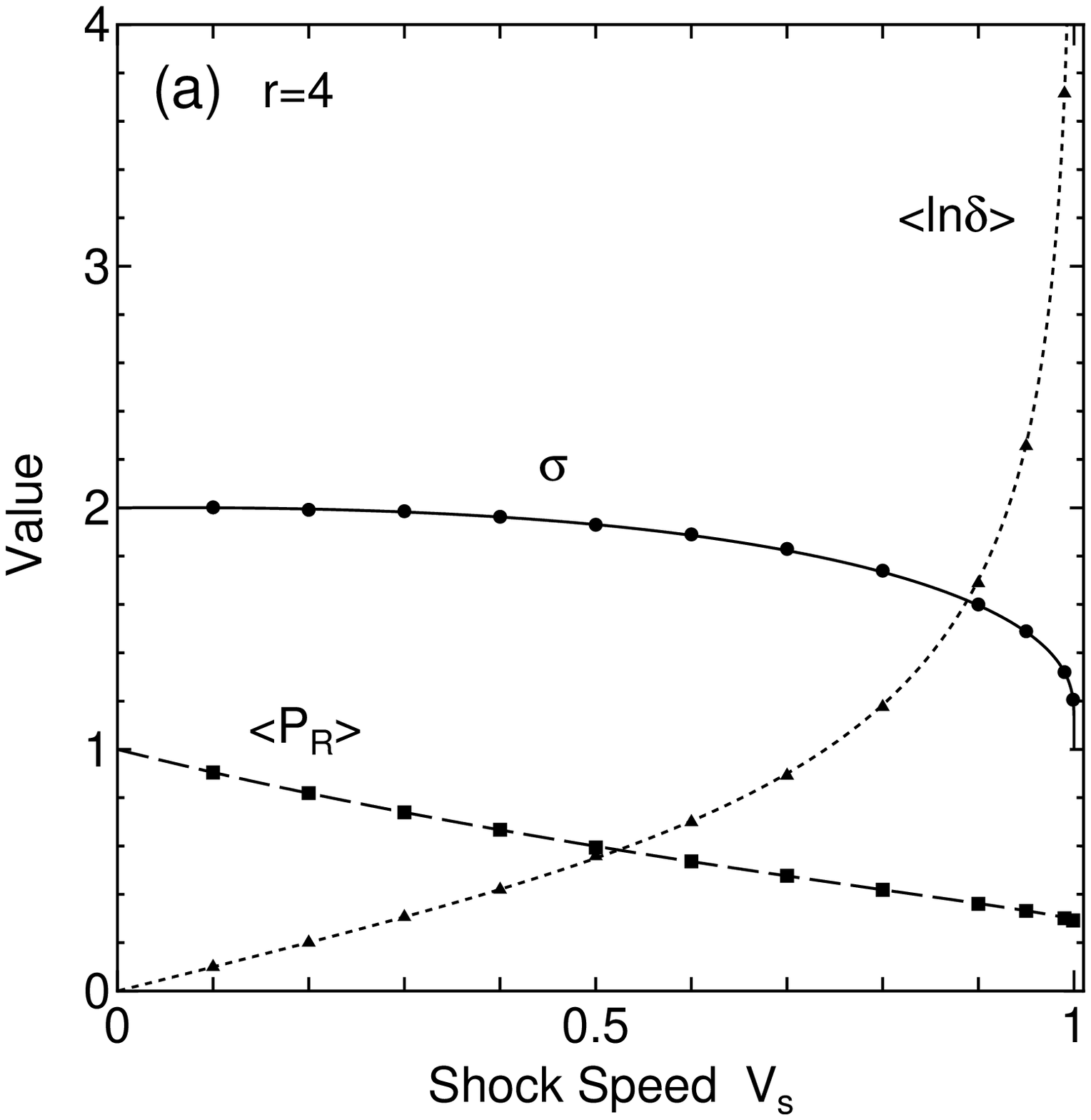}
    \epsfysize=8cm
    \epsfbox{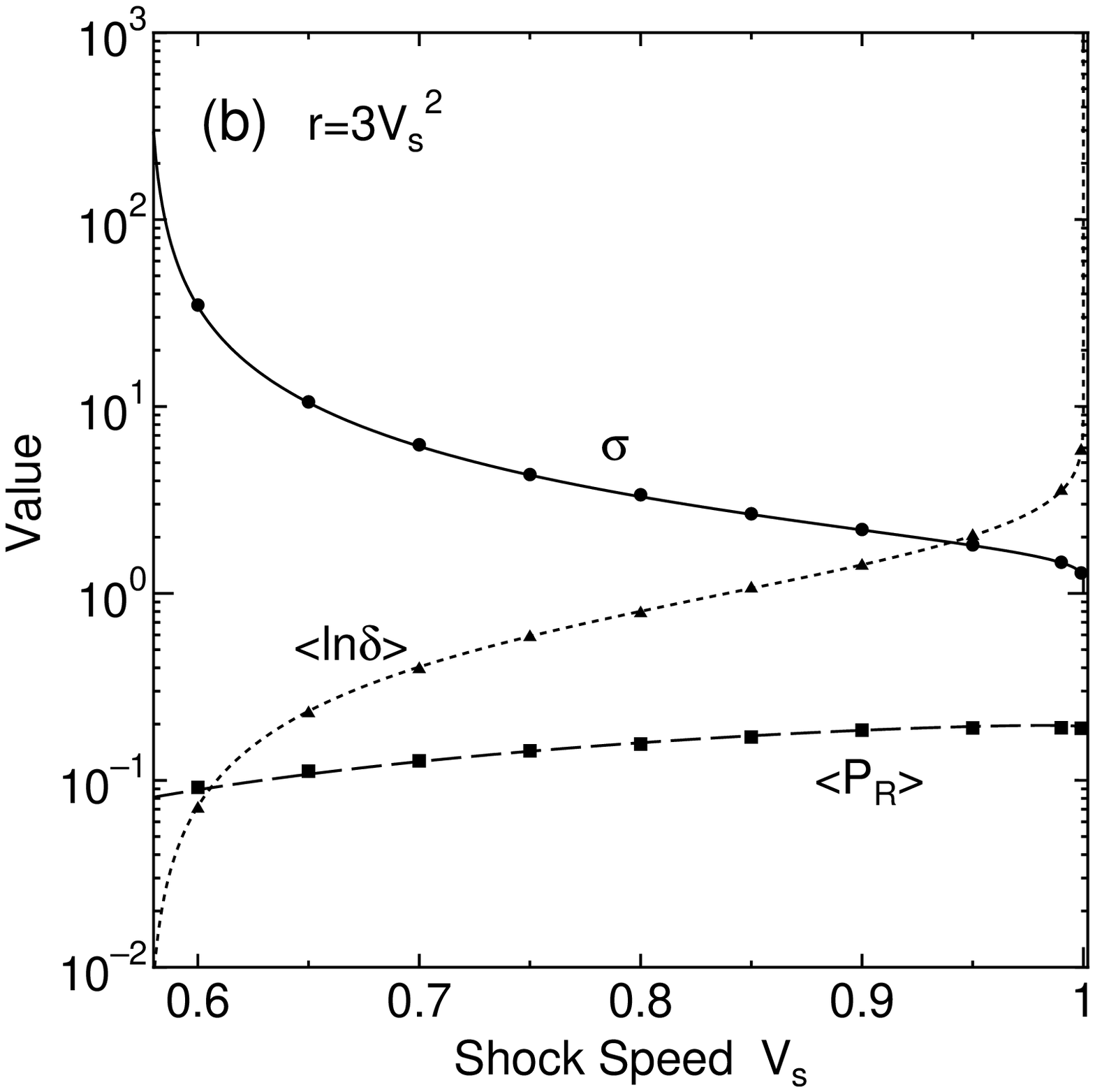}
  \caption{
    The quantities associated with the shock acceleration,  
    the spectral index $\sigma$, return probability $\PRET$ 
    from the downstream and the logarithm of energy gain factor $\lnd$ 
    for the compression ratio $r=4$ (a)
    and for the relativistic jump condition, $\Vu \Vd = 1/3$ (b).
    Curves show the results of the multi-step approximation, 
    while filled symbols show those of 
    Monte Carlo simulation. 
  }
  \label{fig:spectrum}
\end{figure*}
In the previous subsection,
we have shown that
the pitch angle distributions at the shock crossing 
$\pud(\mu)$ and $\pdu(\mu)$
can be well approximated by the multi-step approximation.
Here, using this approximation,
we show analytical expressions of the quantities associated 
with the shock acceleration, $\PRET$, $\lnd$ and $\sigma$,
for the isotropic large angle scattering. 
First, using $\Pret(\mu_0;\dd)$ (\Refeq{approx_Pret_m})
and $\pud(\mu)$ by the multi-step approximation,
$\PRET$ is calculated as
\begin{equation}
\label{eq:PRET_1}
\PRET = A^{\rm (d)} \cdot U(\Vu, \Vd),
\end{equation}
where 
\begin{equation}
U(\Vu, \Vd)
:=
1 + \frac{\chid - 1}{\chiu - \chid} 
\left( 1 - \frac{\psi(\Rc)}{\psi(\Rpu)} \right),
\end{equation}
with
\begin{equation}
\chiu := \frac{1}{\Gu^2 (\Vu - \Vd) \LDu},
\end{equation}
\begin{equation}
\chid := 1 + \frac{1}{\Gd^2 (\Vu - \Vd) \LDd},
\end{equation}
\begin{equation}
\chiz := \frac{\Vu - \Vd}{1 - \Vu},
\qquad
\chione := \frac{\Vu - \Vd}{1 + \Vd},
\end{equation}
\begin{equation}
\Rc := \chiz \chid \qquad (\Rpu = \chiz \chiu).
\end{equation}
$\psi$ is defined in \refeq{psi}.

Then,
the logarithm of the energy gain factor is given by 
\begin{eqnarray}
\lndud
=&
\ln(1 + \Vrel) \nonumber \\
+&
\frac{  \left( \frac{1}{\chid} - \frac{1}{\chiu} \right) \psi(\chiz)
      + (\chid-1) \Psi(\chid) - (\chiu-1) \Psi(\chiu)  }
     {(\chid - 1)\psi(\Rc) - (\chiu - 1) \psi(\Rpu)}
\end{eqnarray}
and 
\begin{equation}
\lnddu
= \ln(1 + \Vrel)
+ \frac{1}{2} \psi(\chione) - \frac{1}{2},
\end{equation}
where 
\begin{equation}
\Psi(x)
:=
\frac{ \ln(1+\chiz) \ln(x - 1) + \dilog{\frac{1}{1-x}} 
- \dilog{\frac{1+ \chiz x}{1 - x}} }
       { \frac{1}{2} (\chiz x)^2 },
\end{equation}
with Dilogarithm $\dilog{x}$ defined by 
(see Abramowitz \& Stegun 1965)
\begin{equation}
\dilog{x} := - \int_{1}^{x} \frac{\ln t}{t - 1} dt.
\end{equation}

Using these results, \refeq{index} gives
the spectral index $\sigma$.
In \Reffig{spectrum} (a) are presented the quantities
associated with the shock accelerations
as functions of shock speed $\Vs$
for the compression ratio $r=4$.
In \Reffig{spectrum} (b) are shown the results for different 
jump condition  ($\Vu\Vd = 1/3$). 
As is seen in these two figures,
the result of the multi-step approximation
excellently agree with the results of the Monte Carlo simulation
even if shock speed becomes highly relativistic.

\begin{figure}
  \leavevmode
    \epsfysize=8cm
    \epsfbox{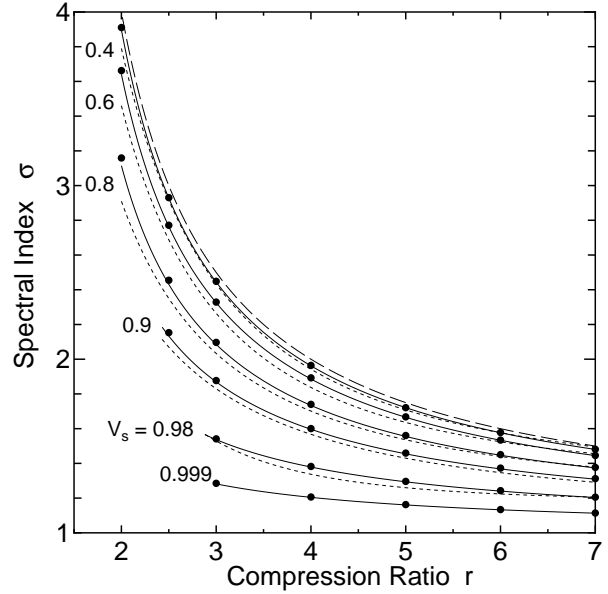}
  \caption{
    The spectral index as functions of the compression ratio $r$
    for various shock speed $\Vs$.
    The solid curves are our results based on the multi-step approximation,
    the dotted curves are the results of fitting formula
    of Ellison et al. (1990)
    and the filled circles are our Monte Carlo results.
    The dashed curve is the results of
    the conventional diffusion approximation,
    i.e., $\sigma = (r+2)/(r-1)$.
  }
  \label{fig:sigma_vs_r}
\end{figure}

\Reffig{sigma_vs_r} presents the spectral index $\sigma$
as functions of the compression ratio $r$
similar to fig.~4 of \EJR90.
Our results (solid curves) basically coincide with
their approximate expression (dashed curves; their equation(29)).
Small discrepancy between these two results
may be caused by the difference of the method
to determine the spectral index $\sigma$; 
we calculate $\sigma$ using \Refeq{index},
while $\sigma$ was determined
by fitting the calculated particle spectrum directly
in \EJR90.

\section{Discussion}
Although
we have considered the particle acceleration in parallel shocks
with the large angle scattering model,
it is of some interest to discuss
whether our formulation can be extended
to other scattering regime.

Recently,
the acceleration in ultra-relativistic shocks
has attracted some attention in relation to
the ultra-high-energy cosmic rays (UHECR).
However,
in such highly relativistic shocks,
the large angle scattering model adopted here
can not be used safely in a physical sense;
because
the residence time of particles in upstream
is quite short
(it is shorter than the gyro-period of particles),
its point-like feature may lead to unsuitable results
(Bednarz \& Ostrowski 1996).
This is also problematic for the pitch angle diffusion model
based on the quasi-linear theory;
because it assumes the resonance between
the magnetohydrodynamical waves and particle gyro-motion,
it requires time much longer than the gyro-period.
Gallant \& Achterberg (1999)
proposed two deflection mechanisms of particles in upstream;
regular deflection owing to the gyro-motion in uniform magnetic field,
and `direction angle' diffusion by
randomly oriented small magnetic cells,
where the direction angle is defined
as the angle between the particle velocity and the shock normal.
Gallant, Achterberg \& Kirk (1999)
and Kirk et al. (2000) calculated the spectral index
using such direction angle diffusion model.
Bednarz \& Ostrowski (1998)
also calculated the index 
adopting the small angle scattering model,
which is approximately equivalent to the direction angle diffusion model
in a mathematical sense.
These works showed the spectral index $\sigma$ typically becomes $2.2$
for ultra-relativistic shock limit,
while our large angle scattering model gives $\sigma \sim 1$.
This index of 2.2 may be universal
when the direction angle diffusion (or equivalent process)
is proper for the scattering/deflection process
in both upstream and downstream,
though
it is not yet settled whether
the real scattering/deflection process during such quite short time
can be treated well as a diffusion in direction angle.

Because
our probability function based
description of the shock acceleration
described in Section 5.1
needs only two probability functions $\Puw$ and $\Pdw$
to obtain the spectral index,
it can be extended to include
other scattering or deflection mechanisms
if corresponding probability functions can be defined.
For example,
for the model of regular deflection in upstream
the probability function $\Puw$ can be defined and expressed in an analytical form
under the condition
that
the direction of velocity perpendicular to the shock normal
is randomly distributed at the shock crossing from downstream.
Thus,
if the large angle scattering model is adopted in downstream (assuming the Bohm diffusion),
one can calculate the spectral index.

\section{Conclusions}
We have given a new formulation of 
the first order Fermi acceleration in shock waves with 
any shock speed in the point of view of the random walk of 
single particles suffering from large angle scattering.  
First, we have investigated the properties of particle trajectories
in a moving medium. 
We showed that the problem could be formulated in terms 
of the random walk with absorbing boundaries
based on the probability theory 
and derived an integral equation for the density of scattering points. 
By approximately solving it in an analytical form,
we have given approximate analytic expressions for 
the probability density of pitch angle at return for particles 
which cross the shock front at a given pitch angle. 
We have confirmed that our approximate solutions agree with 
the Monte Carlo results quite well for the isotropic scattering. 
We have also shown that 
they are quite different from those based on
the diffusion approximation for relativistic and mildly relativistic shocks.
It is seen that the non-diffusive effects such as
the return after only a few steps of scattering
or the finite mean free path,
which are not included in the diffusion approximation,
are very important. 

Using these results we have applied our formulation to the 
shock acceleration and compared our results with those 
in the literature such as `the relativistic diffusion approximation',
which was developed by \Peacock81,
together with the conventional diffusion approximation.
We have found that the multi-step approximation, in which 
the effects of the return after only a few steps of scattering 
are neglected, gives a quite good fit to the Monte Carlo results 
on the pitch angle distributions at the shock crossing and 
the spectral index of accelerated particles. 
This is somewhat surprising because the multi-step approximation 
gives a poor fit to the Monte Carlo results when we fix the 
initial pitch angle at the shock crossing. 
One reason for this is considered that when averaging over 
the initial pitch angle,
the effects of a few steps of return
and the finite mean free path
tend to cancel out and as a result the diffusive return term becomes dominant.  
Thus,
our approach gives a theoretical base to use
the multi-step approximation in the shock acceleration with any shock speed. 
We have also given a correct expression for the diffusive length 
scale for relativistic shocks,
which equals to the scale lengths previously derived by \Peacock81 and
Kirk \& Schneider (1988)
for far upstream distribution in relativistic shocks,
instead of a naive diffusion length 
based on the conventional diffusion equation.

\section*{Acknowledgments}
This work is supported in part by the Grant-in-Aid for Scientific 
Research of the Ministry of Education, Science, Sports and Culture 
(No. 11640236).



\appendix
\section{The relation between the density of scattering points 
and the physical number density}
In order to consider
the connection of the density of scattering points
in random walk to the physical number density,
let us introduce the concept of {\it the staying time density}.
We define the staying time density $S(x) dx$
as the total staying time in the region between $x$ and $x + dx$
for the particle which is injected at the origin $x=0$
and absorbed eventually by one of the absorbing barriers.
The total staying time in the non-absorbing region
before absorption is clearly given by
\[
\int_{-b}^{a} S(x) dx.
\]
Then,
the particle-averaged staying time density $\overline{S}(x)$ 
(i.e., the ensemble average of $S(x)$)
is given as follows.
Consider the contribution to the staying time of a particle 
between the successive scatterings. 
Since the probability of the particles to move beyond $\Dx$ 
without scattering is ${\rm exp}(-|\Dx|/\lmd(\mu'))$
and since the staying time density is $|1/ v \mu|$, 
the pitch angle averaged staying time density
made by the particle during one step, $T(\Dx)$,
is given by 
\begin{equation}
\label{eq:generator}
T(\Dx) =
\left\{
\begin{array}{@{\,}ll}
 \displaystyle 
 \int_{-\nu}^{1} \Pmu \frac{1}{v \mu} e^{-\frac{\Dx}{\lmd(\mu')}} d\mu'
 & (\Dx > 0)\\
 \displaystyle
 \int_{-1}^{-\nu} \Pmu |\frac{1}{v \mu}| e^{\frac{\Dx}{\lmd(\mu')}} d\mu'
 & (\Dx < 0) \\
 \end{array}
\right.
\end{equation}
where $v\mu$ is measured in the boundary rest frame.
Using the probability density of the $m$-th scattering point
$f_m(x)$,
the staying time density made by the particles during  
the $m$-th step to $m+1$-th step
is written as
\[
\int_{-b}^{a} T(x-x') f_m(x') dx'.
\]
Therefore,
summing up all steps from $m=0$ to $\infty$,
the total staying time density is written as
\[
\overline{S}(x) = \int_{-b}^{a} T(x-x') n(x') dx + T(x).
\]
If we consider a steady state problem,
in which particles are injected at the origin
at a constant rate
(with pitch angle distribution $\Pmu$),
$\overline{S}(x)$
gives the physical number density
for unit injection rate.
These relations also have been used in the Monte Carlo simulations
(e.g. Naito \& Takahara 1995).

\section{Derivation of an alternative form of the integral equation}
\label{apn:integral_equation}
The problem of the random walk of single particles,
which is discussed in \Refsec{random walk},
is equivalent to a steady state problem 
in which one particle is injected at the origin per one step.
In the latter problem,
$f_m(x)$ means the number density of scattering points
of particles injected before $m$-steps and
$n(x)$  correspond to the number density of scattering points
made by all particles existing that time.
Here,
we consider the integral equation for $n(x)$
in this equivalent steady problem.

Let us consider the random walk of particles in steady state
with no boundary
but with {\it imaginary} boundaries at $x=a$ and $x=-b$.
The number density of scattering points clearly equal to $\nN(x)$.
Then, we can distinguish 
the particles that cross the boundaries at least once (`passed particles') from 
those that never crossed the boundaries (`non-passed particles').
We write
densities of scattering points of the former and latter 
particles as $\nabs(x)$ and $n(x)$, respectively.
This $n(x)$ obviously coincides with the density of scattering points
of the random walk of single particle
with the absorbing barriers at $x=a$ and $x=-b$.
The sum of these two densities should be the solution for 
the no-boundary problem, $\nN(x)$,
\begin{equation}
\label{eq:nN_n_nabs}
\nN(x) = n(x) + \nabs(x).
\end{equation}
Non-passed particles are injected at the origin steadily
at one particle per step.
On the other hand,
passed particles are {\it created}
when non-passed particles cross one of the imaginary boundaries
and this can be regarded as the 
passed particles are injected at the first scattering point
after they first cross the boundary.
If we denote density of the first scattering point as $D(X)$,
this is calculated by the distribution of non-passed particles $n(x)$ as,
\begin{equation}
D(X) = f(X)
       + \int_{-b}^a f(X-x') n(x') dx' \quad (X>a \; \mbox{or} \; X<-b).
\end{equation}
Each passed particle
which is injected at each first scattering point $X_{\rm inj}$
makes the density of scattering point $n_N(x-X_{\rm inj})$.
Thus, 
$\nabs(x)$ is written as
\begin{eqnarray}
\label{eq:nabs}
\nabs(x)
&=&  \int_a^{\infty} n_N(x-X) D(X) dX \nonumber \\
&&+  \int_{-\infty}^{-b} n_N(x-X) D(X) dX.
\end{eqnarray}
Using \Refeq{nN_n_nabs} and \refeq{nabs},
we obtain
\begin{eqnarray}
n(x)
&=&  n_N(x) - \int_a^{\infty} n_N(x-X) D(X) dX \nonumber \\
&&-  \int_{-\infty}^{-b} n_N(x-X) D(X) dX.
\end{eqnarray}
By the definition of $D(X)$,
\begin{eqnarray}
\int_a^{\infty} n_N(x-X) D(X) dX
=  \int_a^\infty n_N(x-X) f(X) dX \nonumber \\
 + \int_{-b}^a n(x') \int_a^{\infty} n_N(x-X) f(X-x') dX dx',
\end{eqnarray}
\begin{eqnarray}
\int_{-\infty}^{-b} n_N(x-X) D(X) dX
=  \int_{-\infty}^{-b} n_N(x-X) f(X) dX \nonumber \\
 + \int_{-b}^a n(x') \int_{-\infty}^{-b} n_N(x-X) f(X-x') dX dx'.
\end{eqnarray}
Thus,
defining the kernel as
\begin{eqnarray}
k(x,x')
&:=&  \int_a^\infty n_N(x-X) f(X-x') dX \nonumber \\
&&+       \int_{-\infty}^{-b} n_N(x-X) f(X-x') dX,
\end{eqnarray}
we get a Fredholm integral equation of the second kind,
\begin{equation}
n(x) = n_N(x) - k(x,0) - \int_{-b}^a k(x,x') n(x') dx'.
\end{equation}
This provides an alternative form of the integral equation for $n(x)$.

\section{Derivation of integral condition for $\Np(x; \mu'_0)$}
\label{apn:condition}
This is the case of $a \rightarrow \infty$.
We first rewrite the Wald's identity \refeq{a_inf} as
\[
\int_{-\infty}^{-b} e^{-\th0 X}
\int_{-b}^{\infty} f(X-x') \left\{ n(x';\infty,b) + \delta(x') \right\} dx' dX
= 1.
\]
This equation can be rewritten further as
\begin{eqnarray}
  \int_{-\infty}^{0}e^{-\th0 X} \int_{0}^{\infty}
  f(X-x') \left\{ n(x'-b;\infty,b) + \delta(x'-b) \right\} dx' dX
    \nonumber \\
  = e^{-\th0 b}.&&
\end{eqnarray}
On the other hand,
$\Np(x;\mu'_0)$ in the \Refeq{Np}
can be rewritten as
\[
\Np(x;\mu'_0)
= \frac{1}{\lmdz}
  \int_{0}^{\infty} \left\{ n(x'-b;\infty,b) + \delta(x'-b) \right\} e^{-\frac{b}{\lmdz}} db.
\]
Thus,
combining these two,
we obtain the condition to be satisfied for $\Np(x;\mu'_0)$:
\begin{equation}
\int_{-\infty}^{0} e^{-\th0 X}
\int_{0}^{\infty} f(X-x') \Np(x'; \mu'_0) dx' dX
= \frac{1}{1 + \th0 \lmdz}.
\end{equation}
Using the representation of p.d.f. for large angle scattering model \refeq{pdf},
this condition becomes \Refeq{Np_condition}.

\section{The absorption probability by our approximation}
\label{apn:Pabs}
The absorption probability from initial position $b$, $\Pabs(b)$,
 by our approximation is calculated as follows.
\begin{equation}
\Pabs(b) = A' e^{\frac{b}{\LD}} + \eta(b)
\end{equation}
where we define functions
\begin{eqnarray}
\eta(b) &:=&
\frac{n_0 \LD^2}{2} \left\{ (A\cdot(\frac{1}{n_0 \LD}+ 1 + \rho)+\rho) \xi(b) \right. \nonumber \\
&& + (\frac{1}{n_0 \LD} + A +\rho) \Rm E_2(\frac{b}{\Lm}) + \Rm^2 E_3(\frac{b}{\Lm}) \nonumber \\
&& \left.+ \rho A \cdot (\frac{\Rm}{1 - \Rm}) e^{-\frac{b}{\LD}}
\cdot E_2(\frac{b}{\Lm} - \frac{b}{\LD}) \right\},
\end{eqnarray}
\begin{equation}
\xi(b) := E_1(\frac{b}{\Lm}) - e^{-\frac{b}{\LD}} E_1(\frac{b}{\Lm} - \frac{b}{\LD}).
\end{equation}
and a constant
\begin{equation}
A' := n_0' (\hm + \gp(\LD) \LD) + A \left\{ 1 - \frac{n_0'}{2} \frac{\Lm^2}{1-\Rm} \right\}.
\end{equation}

\end{document}